\documentclass[pmlr,twocolumn,10pt]{jmlr} 

\usepackage{booktabs}
\usepackage{siunitx}

\usepackage[switch]{lineno}

\usepackage{hyperref}       
\usepackage{url}            
\usepackage{xspace}
\usepackage{makecell}
\usepackage{adjustbox}
\usepackage{multirow}
\usepackage{mathtools}
\usepackage[T1]{fontenc}
\usepackage{tablefootnote}
\usepackage{titletoc}

\newcommand{\mname}{EHRXDiff\xspace}
\newcommand{\mnamenull}{$\text{EHRXDiff}_{w\_null}$\xspace}

\def\x{{\mathbf x}}

\def\onedot{\ifx\@let@token.\else.\null\fi\xspace}
\def\eg{{\emph{e.g}\onedot}} 
\def\ie{{\emph{i.e}\onedot}}

\def\x{{\mathbf x}}

\theorembodyfont{\upshape}
\theoremheaderfont{\scshape}
\theorempostheader{:}
\theoremsep{\newline}

\jmlrvolume{287}
\jmlryear{2025}
\jmlrsubmitted{} 
\jmlrpublished{} 
\jmlrworkshop{Conference on Health, Inference, and Learning (CHIL) 2025} 

\title[Towards Predicting Temporal Changes in Chest X-rays from EHR Data]{Towards Predicting Temporal Changes in a Patient's \titlebreak Chest X-ray Images based on Electronic Health Records}

\author{%
\Name{Daeun Kyung}\Email{kyungdaeun@kaist.ac.kr}\\
\addr KAIST, Republic of Korea
\AND
\Name{Junu Kim}\Email{kjune0322@kaist.ac.kr}\\
\addr KAIST, Republic of Korea
\AND
\Name{Tackeun Kim}\Email{tackeun.kim@taloscorp.io}\\
\addr TALOS Corp, Republic of Korea
\AND
\Name{Edward Choi\nametag{\thanks{Corresponding author}}} \Email{edwardchoi@kaist.ac.kr}\\
\addr KAIST, Republic of Korea
}

\begin{document}

\maketitle

\begin{abstract}
Chest X-ray (CXR) is an important diagnostic tool widely used in hospitals to assess patient conditions and monitor changes over time.
Recently, generative models, specifically diffusion-based models, have shown promise in generating realistic synthetic CXRs.
However, these models mainly focus on conditional generation using single-time-point data, \ie, generating CXRs conditioned on their corresponding reports from a specific time. This limits their clinical utility, particularly for capturing temporal changes.
To address this limitation, we propose a novel framework, \mname, which predicts future CXR images by integrating previous CXRs with subsequent medical events, \eg, prescriptions, lab measures, etc.
Our framework dynamically tracks and predicts disease progression based on a latent diffusion model, conditioned on the previous CXR image and a history of medical events.
We comprehensively evaluate the performance of our framework across three key aspects, including clinical consistency, demographic consistency, and visual realism. Results show that our framework generates high-quality, realistic future images that effectively capture potential temporal changes. 
This suggests that our framework could be further developed to support clinical decision-making and provide valuable insights for patient monitoring and treatment planning in the medical field.
\end{abstract}

\paragraph*{Data and Code Availability}
This paper uses the MIMIC-IV~\citep{johnson2023mimiciv} and MIMIC-CXR-JPG~\citep{johnson2019mimiccxr} dataset, both of which are available on the PhysioNet repository~\citep{Johnson2023mimic2.2, Johnson2019mimiccxrjpg}. Our implementation code is available at this repository\footnote{\url{https://github.com/dek924/EHRXDiff}}.

\paragraph*{Institutional Review Board (IRB)}
This research does not require IRB approval.

\section{Introduction}
\label{sec:intro}
Chest X-ray (CXR) is a crucial diagnostic tool employed in hospitals to evaluate a patient's health status and monitor changes over time. 
Due to its cost efficiency and low radiation dose, CXR is the most frequently performed imaging technique in hospitals.
This widespread use makes building large-scale datasets relatively easy compared to other medical imaging methods, leading many studies to focus on training AI models using these X-ray datasets.

The recent success of high-quality image generation in the general domain has inspired efforts to create realistic synthetic CXRs in the medical field.
Initial studies~\citep{Loey2020covidgan, sundaram2021ganaug, lee2024ViewXGen} used Generative Adversarial Networks (GANs)\citep{Goodfellow2014GAN} or VQ-VAE\citep{van2017VQGAN}, which were state-of-the-art (SOTA) models for image generation at the time. Recently, diffusion-based models such as RoentGAN~\citep{chambon2022roentgen} and Cheff~\citep{weber2023cheff} have been explored to generate high-fidelity medical images.

These models show promising results in generating CXRs that depict specific diseases based on their labels~\citep{Loey2020covidgan, sundaram2021ganaug}, or modeling desired pathology locations and severities using text descriptions, such as paired CXR reports~\citep{lee2024ViewXGen, chambon2022roentgen, weber2023cheff, Shentu2024CXRIRGen, chambon2022adapting}.
However, these methods are limited to synthesizing realistic CXRs at a single time point and cannot provide future images of specific patients. Thus, they can only serve as data augmentation tools for addressing class imbalances, without leveraging CXRs' ability to track disease progression over time. 
Given that CXR imaging is crucial not only for assessing current health status but also for monitoring disease progression, a new framework that incorporates temporal changes is needed.

Electronic Health Records (EHRs) are large-scale, multimodal databases that encompass patients' comprehensive medical histories. CXRs are also included in EHRs as part of the imaging modalities, alongside structured, tabular data such as diagnoses, procedures, and medications. 
In clinical practice, physicians rely on both patient images and EHR tabular data for decision-making~\citep{Huang2020ehrcxr, Kerchberger2020ehrcxr}. Therefore, it is both natural and potentially impactful to combine prior CXRs with subsequent medical events from the EHR tabular data to predict changes in a patient’s condition.

In this paper, we introduce a novel task of predicting future CXRs by integrating previous CXRs with subsequent medical events from the EHR tabular data.
To tackle this challenge, we propose a latent diffusion-based framework, named \mname.
This approach offers a dynamic view of a patient's temporal changes, tracking their condition based on medical events such as medication or treatment, starting from the initial status of the patient (\ie, the previous CXR image).
We comprehensively evaluate the quality of the predicted CXRs of our proposed model. 
Specifically, our framework shows promise in providing valuable insights for patient monitoring, which can help healthcare professionals, as it outperforms baseline methods in tracking patients’ changing states over time.

\section{Related Works}
\label{sec:related_works}
\subsection{Generative Models for CXR Imaging}
Generative models have been widely employed to tackle challenges in CXR imaging, such as class imbalance and label noise. Early efforts, including COVID-GAN~\citep{Loey2020covidgan} and GAN-Aug~\citep{sundaram2021ganaug}, utilized GANs to synthesize CXRs with specific disease labels for data augmentation. While these methods enhanced the representation of underrepresented classes, they often lacked the ability to produce high-fidelity images with anatomically accurate details.

To address these shortcomings, more advanced techniques such as VQ-VAE~\citep{van2017VQGAN} and Latent Diffusion Model (LDM)~\citep{rombach2022ldm} were introduced. Models like ViewXGen~\citep{lee2024ViewXGen} and LLM-CXR~\citep{ lee2024llmcxr}, leveraged VQ-GAN~\citep{esser2020taming} to generate CXRs using image tokens, combining this capability with large language models (LLMs) to further enhance generative performance. In the other hand, LDM based frameworks, such as RoentGAN~\citep{chambon2022roentgen} and Cheff~\citep{weber2023cheff}, achieved remarkable realism in generating high-quality CXRs, enabling conditional synthesis based on disease labels or text descriptions. Despite their success, these models mainly focused on single-time-point image generation, limiting their use in scenarios that require longitudinal or temporal analyses.

\begin{figure}[t!]
\begin{center}
\includegraphics[width=\linewidth]{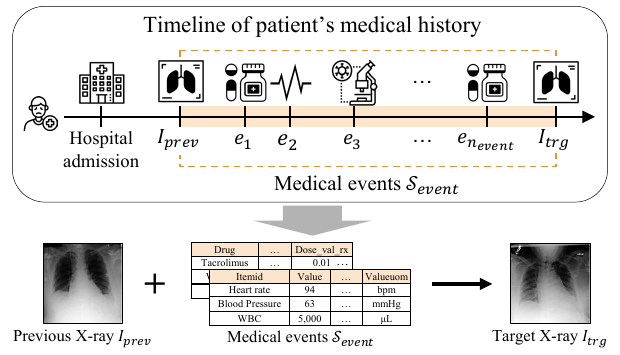}
\end{center}
\vspace{-0.4cm}
\caption{Task overview. EHR containing a patient's medical history occurred within the hospital, including structured data (\eg, charts, medications, microbiology events) and unstructured records (\eg, CXR). Our task is to predict subsequent CXR image based on prior CXR image and the associated medical history after their acquisition.}
\vspace{-0.4cm}
\label{fig:task}
\end{figure}

\subsection{Longitudinal CXR Imaging}
Sequential CXRs provide valuable context for explaining disease progression over time. Accordingly, recent works~\citep{Nicolson2024longitudinalcxr, bannur2024maira2, sharma2024mairaseg} in CXR report generation tasks have utilized prior images from a patient to enhance the quality of generated reports. These studies incorporate patients’ previous studies and corresponding reports as inputs, leveraging temporal information from multiple CXRs to generate detailed clinical narratives. By doing this, they enhance the accuracy of generated reports while simultaneously emphasizing changes in pathological findings over time, demonstrating the value of longitudinal CXR data in multimodal tasks like report generation.

Despite the importance of longitudinal imaging, few studies have focused on the generation or prediction of CXRs over time. 
Recently, BiomedJourney~\citep{gu2023biomedjourney} introduced a counterfactual image generation framework that synthesizes CXRs based on hypothetical disease progression scenarios.
Given two CXR reports taken at different time points for a specific patient, GPT-4~\citep{openai2024gpt4technicalreport} is used to synthesize the disease progression between these CXRs. While effective in generating diverse imaging outcomes, this approach serves a different purpose from ours. They focus on generating CXRs based on specific progression scenarios, rather than predicting the most likely future CXRs based on the actual medical history of patient. Consequently, it does not address the critical need for predicting patient-specific future CXRs by incorporating medical history, leaving a significant gap in leveraging temporal imaging data for generative tasks.

\subsection{Multimodal Fusion for Imaging and EHR Tabular Events in Clinical Applications}
Clinical decision-making often relies on the integration of multiple data modalities~\citep{Huang2020ehrcxr, Kerchberger2020ehrcxr, bae2023ehrxqa}, including imaging and electronic health records (EHRs). multimodal fusion methods, combining CXRs with structured EHR data, have primarily been explored for clinical time-series prediction tasks~\citep{hayat2023medfuse, DrFuse2024Yao}, such as in-hospital mortality prediction and phenotype classification. These studies consistently demonstrate that leveraging both tabular EHR features and imaging data improves model performance for classification and prediction tasks. However, multimodal methods for CXR generative modeling remain under-explored. Predicting future CXRs conditioned on prior imaging and EHR data requires integrating temporal dynamics and multimodal context, a gap that this study aims to address.

\section{Methodology}
\label{sec:method}

\subsection{Task Definition}
Our goal is to predict a target CXR image $I_{trg}$, given a previous CXR image $I_{prev}$ and a sequence of medical events $\mathcal{S}_{event} = (e_1, e_2, \ldots, e_{n_{event}})$ (\figureref{fig:task}).
We work with the triple ($I_{prev}, \mathcal{S}_{event}, I_{trg}$), where $I_{prev}, I_{trg} \in \mathbb{R}^{3 \times H \times W}$, with $H$ and $W$ denoting the input image resolution (height and width), respectively. 
The sequence $\mathcal{S}_{event}$ consists of medical events $e_i$ from the EHR tabular data (\eg, prescriptions, lab tests) occurring at time $t_i$ between the two imaging times. Here, $i \in \{1, \dots, n_{event}\}$, with $n_{event}$ indicating the total number of events. 
This task is more challenging than the previous works~\citep{chambon2022roentgen, weber2023cheff, chambon2022adapting} since it incorporates the patient's medical history rather than relying on simple text descriptions.
The model must recognize the initial state from the previous image and integrate temporal information from the medical event sequence to predict the patient's status in the target image.

\subsection{Background: Conditional Latent Diffusion Model}
We adopt the state-of-the-art image generation model, Latent Diffusion Model (LDM) \citep{rombach2022ldm}, as our backbone. The LDM consists of two main components: (1) A Variational Autoencoder (VAE)~\citep{van2017VQGAN} with encoder $\textit{E}_{\mathrm{VAE}}$ and decoder $\textit{D}_{\mathrm{VAE}}$, which maps an image to a lower-dimensional latent representation $z$ and reconstructs the image from it, and (2) a conditional denoising U-Net $\epsilon_{\theta}$ that iteratively removes noise from a noised latent vector $z_t$.

During training, we first obtain the initial latent representation $z_0 = E_{\mathrm{VAE}}(\mathbf{x})$ for an input image $\mathbf{x}$. This latent $z_0$ is then progressively corrupted by iteratively adding noise according to:
\begin{equation}
    \begin{split}
        z_t &= \sqrt{\bar{\alpha}_t}\,z_0 + \sqrt{1 - \bar{\alpha}_t}\,\epsilon, \\
        \epsilon &\sim \mathcal{N}(0,I), \quad t \in \{1,\dots, T\}.
    \end{split}
\end{equation}
where $\bar{\alpha}_t = \prod_{s=1}^t \alpha_s$, with $\alpha_s = 1 - \beta_s$ for a given variance schedule $\{\beta_s\}_{s=1}^T$.  

The model is trained to predict the noise $\epsilon$, with the objective function defined as:
\begin{equation}
    \mathcal{L}_{\mathrm{LDM}}= \mathbb{E}_{z_t \sim \mathcal{N}(0,1), t} \bigl\| \epsilon_t - \epsilon_{\theta}(z_t, \tau_{\phi}(\textbf{y}), t) \bigr\|_2^2,
\end{equation}
where $\tau_{\phi}$ is an embedding model for the condition $\textbf{y}$. 
The denoising U-Net incorporates the condition embedding either via cross-attention or by concatenating $\tau_{\phi}(\textbf{y})$ with the noised latent $z_t$. 
Following \citet{rombach2022ldm, weber2023cheff}, for cross-attention, the model fuses $\tau_\phi(\textbf{y})$ with its intermediate feature representation $\epsilon_i$ though attention layer $\mathrm{Attention}(Q,K,V) = \mathrm{softmax}\!\Bigl(\tfrac{Q\,K^\top}{\sqrt{d}}\Bigr)\,V$. 
We form the queries, keys, and values as:
\[
    Q = \epsilon_i \cdot W_Q^{(i)},\quad 
    K = \tau_{\phi}(\mathbf{y}) \cdot W_K^{(i)},\quad 
    V = \tau_{\phi}(\mathbf{y}) \cdot W_V^{(i)},
\]
where $\epsilon_i \in \mathbb{R}^{N \times d_{\epsilon}^i}$ denotes a intermediate representation of the U-Net at the $i$-th layer where $N$ is the flattened spatial dimension. $W_Q \in \mathbb{R}^{d_{\epsilon}^i \times d}$ and $W_K, W_V \in \mathbb{R}^{d_{\tau} \times d}$ are learnable parameter matrices.
         
In our setting, we treat a previous CXR image $I_{\mathrm{prev}}$ and a sequence of medical events $\mathcal{S}_{\mathrm{event}}$ as the conditioning $y$. We further detail how $\tau_{\phi}(\mathbf{y})$ is constructed from these inputs and integrated into the U-Net in below section.

\begin{figure*}[t!]
\begin{center}
\includegraphics[width=0.95\textwidth]{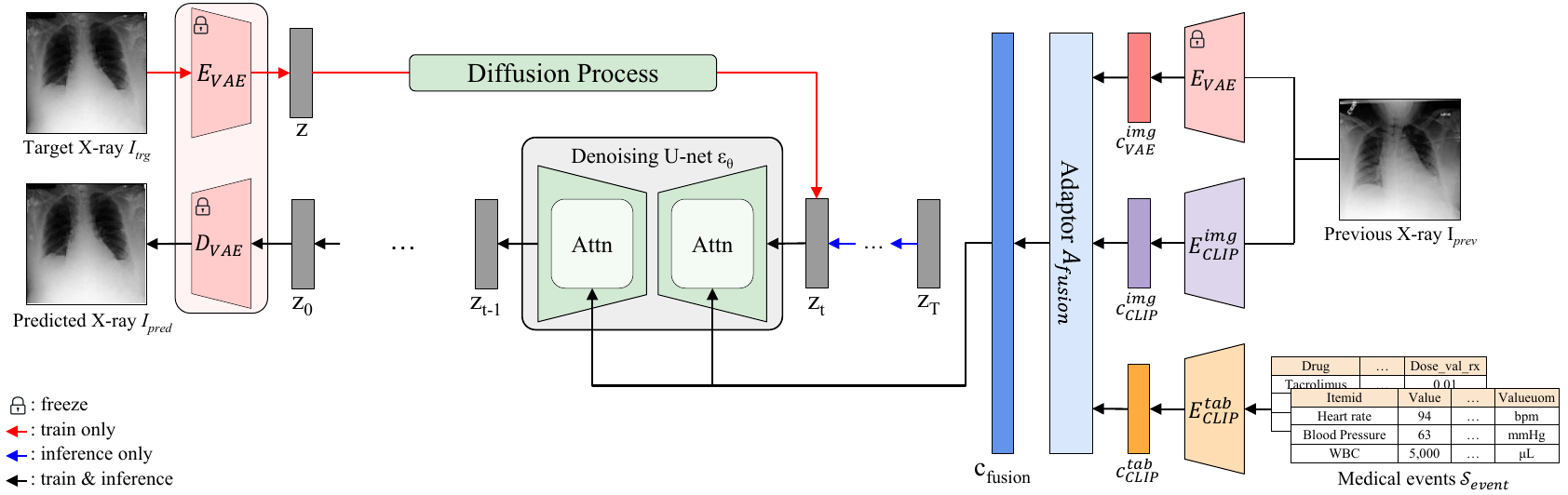}
\end{center}
\vspace{-0.6cm}
\caption{Overall framework. During training, a random timestep $t$ is sampled, and the latent vector $z$ is corrupted to $z_t$ via diffusion process. For inference, Gaussian noise is sampled and iteratively denoised over $T$ steps. In both cases, image embeddings from CLIP and VAE encoders ($E_{CLIP}^{img}$, $E_{\mathrm{VAE}}$) and table embeddings from CLIP table encoder ($E_{CLIP}^{tab}$) are fused by an adaptor module to condition the denoising U-Net.}
\vspace{-0.5cm}
\label{fig:model_overall}
\end{figure*}
\begin{figure}[t!]
\begin{center}
\includegraphics[width=0.88\linewidth]{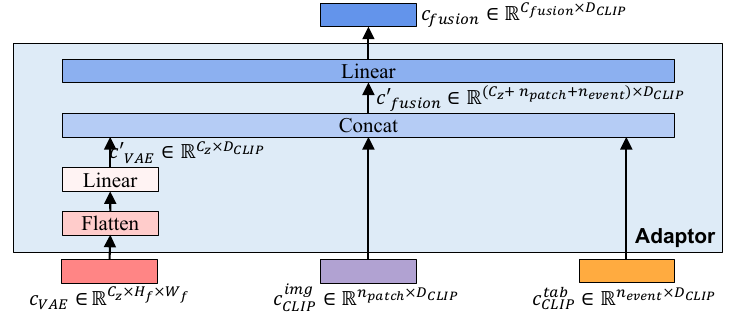}
\end{center}
\vspace{-0.5cm}
\caption{Detail of adapter module $A_{fusion}$. This module merges the embeddings from the VAE and CLIP encoders. To simplify the process, we project the VAE embeddings to match the dimensionality of the CLIP features $D_{CLIP}$ before fusing.}
\vspace{-0.6cm}
\label{fig:adaptor}
\end{figure}

\subsection{Model Architecture}
\label{sec:model_architecture}
To predict future CXR images based on previous CXR images and consecutive medical events, the model must (1) produce realistic, high-quality CXR images and (2) accurately reflect the potential clinical status based on the prior image and medical history. To achieve this, the framework includes three encoders: a VAE encoder, and two CLIP encoders for each of the image and table modalities. 
The outputs from these encoders are then integrated using an adapter module, efficiently leveraging both the previous images and the medical events. Detailed descriptions of each module are provided below.

\subsubsection{Encoder Modules}
\paragraph{VAE Encoder} The VAE encoder, $\textit{E}_{\mathrm{VAE}}$, captures fine-grained visual details such as the shapes of major organs (\eg, the lungs and heart) as well as the the locations and sizes of the lesions. These details are crucial for understanding and preserving the structural information of the input CXR image. We denote the VAE embedding of the previous CXR image as $c_{\mathrm{VAE}}^{img} \in \mathbb{R}^{C_z \times H_f \times W_f}$, where $C_z$ is the channel size of the latent vector, and $H_f$ and $W_f$ represent the spatial resolution of the feature map.

\paragraph{CLIP Encoders} We use separate CLIP encoders for two modalities: the image CLIP encoder ($\textit{E}_{CLIP}^{img}$) and the table CLIP encoder ($\textit{E}_{CLIP}^{tab}$). 
Typically, LDMs use CLIP~\citep{radford2021clip}, which is pretrained on large-scale image-text pairs, as a conditioned embedding model.
Since there are no existing pretrained models for image and tabular modalities in the medical domain (\eg, CXR images and medical events from EHR tabular data), we replace the text encoder in the CLIP model with a table encoder. We then pretrain the image and table encoders on pairs of prior CXR images and their associated medical events ($I_{prev}, \mathcal{S}_{event}$).

The $\textit{E}_{CLIP}^{img}$ extracts high-level clinical features related to the patient's initial clinical status from $I_{prev}$, while the $\textit{E}_{CLIP}^{tab}$ captures relevant information from medical events, effectively interpreting the patient's medical history and its impact on their condition. 
We extract the CLIP embedding $c_{CLIP}^{img} \in \mathbb{R}^{n_{patch} \times D_{CLIP}}$ and $c_{CLIP}^{tab} \in \mathbb{R}^{n_{event} \times D_{CLIP}}$ from the input CXR and medical event sequences, respectively. Here, $n_{patch}$ denotes the number of visual patches from the Vision Transformer (ViT) encoder~\citep{dosovitskiy2020vit}, $n_{event}$ is the number of medical events, and $D_{CLIP}$ is the dimensionality of the CLIP embeddings.

\subsubsection{Adapter for Multimodal Fusion}
In previous works on latent diffusion models that using image condition~\citep{rombach2022ldm, gu2023biomedjourney, brooks2023instructpix2pix}, the conditioning image embedding is often concatenated with the input noise of the denoising U-Net.
This approach effectively preserves spatial alignment and is well suited for tasks such as image-to-image translation or super-resolution.
However, for CXR image pairs, even consecutive images may be misaligned due to factors such as patient posture or breathing variations~\citep{Dasegowda2023Suboptimalcxr}. Thus, our task requires integrating diverse information from prior CXR image rather than simply replicating structural details. 
To achieve this issue, we replace the concatenation strategy with a cross-attention mechanism using an adapter module $\textit{A}_{fusion}$ as described below. This method enables the model to focus on meaningful changes rather than enforcing strict spatial alignment.

The adapter module, $\textit{A}_{fusion}$, merges the embeddings from the VAE and CLIP encoders to fuse the fine-grained structural details from the VAE with the high-level clinical features from the CLIP encoders (\figureref{fig:adaptor}).
Specifically, the VAE embedding, $c_{\mathrm{VAE}}^{img}$, is a 2D feature map per channel, while the CLIP embeddings, $c_{CLIP}^{img}$ and $c_{CLIP}^{tab}$, are 1D vectors per token. To make these embeddings compatible, we first flatten $c_{\mathrm{VAE}}^{img}$ into a 1D format and project it to match the dimensionality of the CLIP features, $D_{CLIP}$. 
Then, the projected VAE features are concatenated with the CLIP features, and the combined embeddings are passed through a linear layer, creating a fused embedding $c_{fusion}$. The resulting embedding $c_{fusion}$ has $C_{fusion}$ channels, each with a dimensionality of $D_{CLIP}$.

This fusion allows the model to predict future CXR images that not only represent the structural and clinical nuances of the input images but also reflect potential temporal changes based on medical events. By doing so, the model leverages both the spatial detail from VAE embedding and the contextual understanding from the CLIP embeddings, ensuring that predictions are both anatomically accurate and clinically relevant.

\subsection{Data Augmentation}
\label{sec:data_aug}
To help the model learn correlations within our triplets, we introduce an additional technique that creates a \textit{no-change} sample $(I_{prev}, \mathcal{S}_{null}, I_{prev}')$ where $I_{prev}$ and $I_{prev}'$ represent the same CXR image with different augmentations, and $\mathcal{S}_{null}$ represents an empty set of medical events.
The augmentation from $I_{prev}$ to $I_{prev}'$ involves weak transformations such as random rotation or scaling, which simulate patient movements during imaging while preserving clinically relevant information. This approach helps the model maintain consistency in clinical findings when there are no significant events in the EHR table, thus teaching it to distinguish between clinically meaningful changes and minor variations.
We present two versions of our model, \mname and \mnamenull, both of which share the same architecture (see \figureref{fig:model_overall}). \mnamenull additionally utilizes the null-based augmentation technique described above.

\section{Experiments}
\subsection{Dataset}
\paragraph{MIMIC-IV (v2.2)}
MIMIC-IV~\citep{johnson2023mimiciv} is a freely accessible relational database with de-identified health data from 50,920 patients admitted to the critical care units at Beth Israel Deaconess Medical Center (BIDMC) between 2008 and 2019. It includes detailed information on demographics, lab results, diagnoses, procedures, medications, and more, supporting diverse clinical research.

\paragraph{MIMIC-CXR-JPG (v2.0.0)} MIMIC-CXR~\citep{johnson2019mimiccxr} is a public dataset of 377,110 chest radiographs from 227,827 imaging studies at BIDMC between 2011 and 2016. We use MIMIC-CXR-JPG~\citep{Johnson2019mimiccxrjpg}, a derivative of MIMIC-CXR, which provides standardized JPG images generated from the original DICOM files. This dataset is the only imaging resource directly linked to MIMIC-IV, enabling robust multimodal research.

\subsection{Data Preprocessing}
\label{sec:dataset}
We extract $(I_{prev}, \mathcal{S}_{event}, I_{trg})$ triples from the MIMIC-IV and MIMIC-CXR-JPG datasets.
For medical events, we select seven core event types: lab, chart, input, output, prescription, procedure, and microbiology. These event types are chosen for their relevance to patient status, as determined in guidance with a medical expert. From CXR images, we include only frontal-view chest X-rays (\ie, PA and AP views).

To integrate medical events with CXR images, we select patients with at least two CXR images taken during hospitalization. We focus on CXR pairs captured within a maximum of two days, aiming to exclude outliers while accommodating input size limitations. This approach is informed by the observation that approximately 80\% of consecutive CXRs are obtained within this time frame.

For the selected CXR pairs, we extract the list of medical events occurring between the two imaging times.
Given the abundance of EHR data during ICU stays, where patients often experience thousands of events per day~\citep{Sanchez2018icu}, we focus on cardiovascular and lung-related items from the selected tables that could affect CXR images. This selection is validated by a medical expert to ensure clinical relevance. Each tabular event $e_i$ is converted into free-text format, such as ``\textit{table\_name column1 value1} $\dots$ ,'' following the methodology of \citet{Hur2024GenHPF}. These free-text representations are then embedded into vector representations using OpenAI's embedding model (\texttt{text-embedding-ada-002}). 

We randomly split the data into a 90:5:5 ratio at the sample level while ensuring that patient cohorts do not overlap across the splits. This resulted in 101,819 samples for training, 5,654 for validation, and 5,596 for testing (\tableref{tab:statistics}). Details of the data preprocessing are provided in \appendixref{apd:data_preprocess}.

\begin{table}[t!]
\vspace{-0.3cm}
\caption{Overall statistics of our dataset.}
    \begin{center}
    {\normalsize{
        \resizebox{1\linewidth}{!}{
            \begin{tabular}{lccc}
            \toprule
                                                      & Train      & Valid    & Test      \\ \midrule
        \# of patients                                & 9,295       & 595      & 407     \\
        \# of hospital admission ids                  & 11,263      & 730      & 523     \\
        \# of samples ($I_{prev}, \mathcal{S}_{event}, I_{trg}$)  & 101,819    & 5,654    & 5,596    \\
        avg. time interval per sample (hours)           & 25.64     & 25.87   & 25.84     \\
        \bottomrule
        \end{tabular}
    }}}
    \end{center}
\label{tab:statistics}
\vspace{-0.5cm}
\end{table}


\subsection{Evaluation Metrics}
We evaluate the quality of the predicted X-rays using three main categories of metrics: (1) the preservation of medical information, ensuring disease details are accurately represented, (2) the preservation of demographic information (\ie, age, gender, race), maintaining key patient attributes from the input image, and (3) overall image quality, ensuring realistic, high-quality chest radiographs.

\subsubsection{Preservation of Medical Information}
\label{section:evaluation_cls}
We evaluate medical information preservation by performing multi-label classification for CXR pathologies, based on labels from Chest ImaGenome~\citep{wu2chestImaGenome} database and extracted using CheXpert labeler~\citep{irvin2019chexpert}. 
Since Chest ImaGenome provides a broader range of CXR diagnoses, we mainly focus on its labels.
For Chest ImaGenome, we fine-tune a pretrained Vision Transformer (ViT) model on the MIMIC-CXR-JPG dataset, following \citet{bae2023ehrxqa}, using our training set. To ensure robust evaluation, we focus on the 10 most frequent Chest ImaGenome labels, each with a prevalence between 0.05 and 0.95. We report weighted macro-average AUROC across these diagnoses, which helps prevent skewed assessments and enhances the reliability of performance metrics.
Details on the CheXpert-based evaluation settings are provided in the \appendixref{apd:chexpert_implementation}.

We categorize results into three groups: 
\begin{itemize} 
    \item \textit{All}: The overall performance on the entire test set. 
    \item \textit{Same}: The subset where the previous and target labels are identical. 
    \item \textit{Diff}: The subset where the previous and target labels are different. 
\end{itemize} 
We adopt these three groups because it is crucial to (1) preserve disease information when there is no change (\textit{Same}) and (2) capture temporal changes accurately when they do occur (\textit{Diff}). 
Further details about the Chest ImaGenome settings are provided in \appendixref{apd:chestimagenome_implementation}.

\subsubsection{Preservation of Demographic Information}
To assess the preservation of demographic information, we employ classifiers for age, gender and race, using information from the admissions and patients tables in the MIMIC-IV database as reference labels.
For age and race, we utilize an SOTA prediction model sourced from the XRV collection~\citep{Ieki2022age, Gichoya2022race}, following \citet{gu2023biomedjourney}.
Given that most MIMIC patient data spans a short time period, we assume that two images for a given patient correspond approximately to the anchor age provided in the MIMIC database, following \citet{gu2023biomedjourney}. 
For gender, we fine-tune the XRV DenseNet-121 classifier~\citep{cohen2020limits}. We report the AUROC for gender and race predictions and the Pearson correlation for age predictions. More details are provide in \appendixref{apd:demographic_cls}.

\subsubsection{Image Quality} 
We calculate the Fr\'{e}chet Inception Distance (FID) score~\citep{Heusel2017fid} to compare feature statistics between the original CXR images from the test set and the predicted CXR images. Features are extracted using a 1024-dimensional DenseNet-121 model trained on multiple CXR datasets~\citep{cohen2020limits}, as part of the XRV collection~\citep{Cohen2022xrv}.


\subsection{Baselines}
\label{sec:baseline}
As our work is the first to predict temporal changes in CXR images using both previous CXR images and medical events, there are no existing baseline models for direct comparison. Therefore, we define four baselines: \textit{Previous image}, \textit{Previous label}, \textit{Table classifier}, and \textit{Table classifier (w/ prev label)}.

\textbf{\textit{Previous image}}
This baseline directly uses the previous CXR image $I_{prev}$ as future prediction, representing a scenario in which no temporal change is predicted. This approach can act as a potential shortcut for the future CXR prediction model, because simply copying the input previous image is much easier than learning changes driven by complex medical events. 

If the patient’s condition truly remains unchanged (\textit{Same} in \sectionref{section:evaluation_cls}), using the previous image may yield near-perfect pathology classification results, making it an upper bound on performance for this subset.  
However, if a patient’s condition changes (\textit{Diff}), using the previous image will lead to inaccurate pathology classification for the new findings, making it a lower bound for this subset.
Note that we evaluate the preservation of medical information in the predicted CXR by processing it through a separate diagnostic classifier. Due to the classifier’s inherent uncertainty, the extracted labels do not map exactly to 0 or 1, despite the GT labels being binary.

\textbf{\textit{Previous label}} 
Instead of using the previous image, this baseline directly compares the GT labels of $I_{prev}$ and $I_{trg}$.
It represents the same ``no change'' scenario as \textit{Previous image}, but without involving the uncertainty of an image-based classifier in the evaluation step. Thus, it can be viewed as a more strict version of the \textit{Previous image} baseline.
For example, if the pathology labels do not change, this baseline will perfectly match the target’s labels, acting as a strict upper bound for unchanged cases (100\% accuracy). In contrast, if pathology changes, this method will fail to capture the new label, serving as a lower bound in such cases (0\% accuracy).

\textbf{\textit{Table classifier}} 
This baseline predicts CXR diagnosis labels directly from medical event data, without the need to generate or interpret CXR images. We implement this approach by fine-tuning our CLIP table encoder ($\textit{E}_{CLIP}^{tab}$) and adding a linear multi-label classification layer on top. In doing so, we gain a straightforward measurement of how well a model can perform using only structured data, without the complexity of image synthesis.

\textbf{\textit{Table classifier (w/ prev label)}} 
In this extended version, we incorporate the GT label from the previous CXR image by converting it into a learnable label embedding, which is then concatenated with the tabular features. Beyond this addition, the architecture and training procedures remain the same as in the \textit{Table classifier}. By integrating prior label information, the model can more directly leverage a patient’s initial status.
This baseline thus evaluates how much performance improves when the model receives an explicit hint about the patient’s past condition instead of leveraging the actual image.

\subsection{Implementation Details}
\label{sec:implement_detail}
We implement our LDM model based on \citet{weber2023cheff}. For the VAE encoder, we utilize a pretrained model from MaCheX~\citep{weber2023cheff}. The CLIP encoders consist of ViT-B/32~\citep{dosovitskiy2020vit} for images and a 2-layer Transformer encoder for tabular data. The table Transformer is configured with a maximum input length of 1024, a feature dimension of 1536, and 24 attention heads. The input image resolution, $H \times W$, is set to $256 \times 256$. We set $C_z$, $H_f$, and $W_f$ to 3, 64, and 64, respectively. The CLIP feature dimension $D_{CLIP}$ is 768, and $n_{patch}$ is 65. We set $C_{fusion}$ to 1024. 
Each model is trained for 100 epochs using a batch size of 128 with the AdamW optimizer~\citep{loshchilov2019adamw}, initialized with a learning rate of 5e-5. All models are implemented in PyTorch~\citep{pytorch} and trained on an RTX A6000 GPU. Due to the stochastic nature of diffusion model inference, we generate three predictions using different random seeds and report the mean and standard deviation for each metric. Additional details are provided in \appendixref{apd:implementation_detail}.

\begin{table}[t!]
  \caption{
  Quantitative results on the test set are grouped into \textit{All}, \textit{Same}, and \textit{Diff} categories, based on Chest ImaGenome labels. \textit{All} reflects the overall performance, while \textit{Same} and \textit{Diff} correspond to subsets where the labels for $I_{prev}$ and $I_{trg}$ are identical or different, respectively. We report the macro-weighted average AUROC. For our proposed model, we report the mean $\pm$ standard deviation across three seeds.
  }
  \vspace{-0.5cm}
  \begin{center}
    {\small{
    \resizebox{\linewidth}{!}{%
        \begin{tabular}{l|*{3}{>{\centering\arraybackslash}p{1.5cm}}}
            \toprule
                                        & \textit{All} & \textit{Same} & \textit{Diff}  \\
            \midrule            
            GT                          & 0.842 & 0.905 & 0.722   \\  
            \textit{Previous image}              & 0.798 & 0.902 & 0.558   \\  
            \textit{Previous label}              & 0.672 & 1.000 & 0.000   \\  
            \midrule
            \textit{Table classifier}                 & 0.624 & 0.682 & 0.502    \\  
            \textit{Table classifier (w/ prev label)} & 0.743 & 0.984 & 0.116     \\  
            \midrule
            EHRXDiff (Ours)                      & $0.723_{ \pm 0.003}$ & $0.796_{ \pm 0.003}$ & $0.576_{ \pm 0.005}$  \\  
            $\text{EHRXDiff}_{w\_null}$ (Ours)   & $0.764_{ \pm 0.004}$ & $0.844_{ \pm 0.004}$ & $0.580_{ \pm 0.008}$  \\  
            \bottomrule
        \end{tabular}
    }}}
  \end{center}
  \label{table:test_cls}
    \vspace{-0.1cm}
    \footnotesize $^\ast$ The \textit{Previous image} and \textit{Previous label} baselines both serve as upper bounds for the \textit{Same} subset and as lower bounds for the \textit{Diff} subset.
  \vspace{-0.2cm}
\end{table}

\begin{table}[t!]
    \caption{Quantitative results for image quality and demographic preservation on the test set (mean $\pm$ std). ($\uparrow$) denotes the higher score is better and ($\downarrow$) denotes the lower score is better.}
    \vspace{-0.6cm}
    \begin{center}
    {\small{
    \resizebox{1\linewidth}{!}{%
        \begin{tabular}{l|c|ccc}
            \toprule
            & FID (\(\downarrow\)) & \makecell{Age (\(\uparrow\))\\(Pearson corr.)} & \makecell{Gender (\(\uparrow\))\\(AUROC)} & \makecell{Race (\(\uparrow\))\\(AUROC)} \\ 
            \midrule             
            GT                          &    -   &   0.719   &   0.995  & 0.969 \\  
            \textit{Previous image}              &  0.16  &   0.740   &   0.996  & 0.967 \\  
            \midrule
            EHRXDiff                        & $6.08_{ \pm 0.15}$  &  $0.453_{ \pm 0.007}$  & $0.963_{ \pm 0.002}$ & $0.807_{ \pm 0.002}$ \\  
            $\text{EHRXDiff}_{w\_null}$     & $4.57_{ \pm 0.07}$  &  $0.554_{ \pm 0.012}$  & $0.979_{ \pm 0.001}$ & $0.826_{ \pm 0.002}$ \\  
            \bottomrule
        \end{tabular}
    }}}
    \end{center}
    \label{table:test_demo}
    \vspace{-0.5cm}
\end{table}


\section{Results}
In this section, we provide a comprehensive evaluation of our proposed models, \mname and \mnamenull, in comparison with the four baselines defined in \sectionref{sec:baseline}. Both models share the same architecture, with \mnamenull utilizing an additional augmentation technique described in \sectionref{sec:data_aug}.

\subsection{Preservation of Medical Information} 
We evaluate our model's performance in preserving medical information using labels from Chest ImaGenome and CheXpert. Since the classifier's performance on the GT labels of Chest ImaGenome (0.842) is higher than that of CheXpert (0.802) and covers a wider range of diagnoses, our analysis mainly focuses on Chest ImaGenome in this section. Results for CheXpert labels are provided in \appendixref{apd:chexpert_result}.

Overall, our approach demonstrates consistently strong performance across all three evaluation groups for Chest ImaGenome labels (\tableref{table:test_cls}). Below, we present an analysis of each group, with more detailed per-label performance results available in \tableref{table:supp_chestimagenome}.

\textbf{\textit{All}}: 
For the entire test set (\textit{All}), both of our proposed models (\mname and \mnamenull) outperform the table classifier baselines, even though those baselines are designed for multi-label classification without the added complexity of generating a CXR image. 
This result underscores the difficulty of accurately predicting a patient's future state solely from their medical history (\textit{Table classifier}). 
Even when the patient’s previous label information is included (\textit{Table classifier (w/ prev label)}), the model tends to overfit to the “no change” scenario rather than effectively extracting important features from the complex medical history. This overfitting likely occurs because label-based representations are simpler to interpret than image embeddings.

As described in Section~\ref{sec:baseline}, the \textit{Previous image} and \textit{Previous label} baselines both assume ``no change'' in a patient’s condition. This assumption serves as an upper bound for the \textit{Same} subset and a lower bound for the \textit{Diff} subset.
Given that \textit{Same} accounts for 70\% of the test set, these baselines show relatively high overall performance, making a direct comparison somewhat unfair.
Still, our approach remains competitive with \textit{Previous image} and surpasses \textit{Previous label}. This is reasonable because \textit{Previous label} is a stricter baseline, producing zero scores whenever there is a change, whereas \textit{Previous image} can yield partially correct predictions due to the inherent uncertainty of the diagnostic classifier.

\textbf{\textit{Same}}: 
When the pathology labels remain unchanged between the previous and target CXR images, both of our proposed models achieve high AUROC, comparable to the upper bound performance (\textit{Previous image}). Specifically, \mnamenull outperforms \mname due to its augmentation technique, which helps it avoid modifying clinical findings when no significant medical events occur.
As a result, \mnamenull reaches a 94\% AUROC in the \textit{Same} subset compared to the GT performance. This demonstrates its ability to detect the initial status of the given prior CXR image while accurately recognizing when the patient’s status remains unchanged.
Compared to table classifier variants, our methods outperform the \textit{Table classifier} but underperform relative to \textit{Table classifier (w/ prev label)}.
However, this gap is expected, as the latter can simply copy the previous label, guaranteeing near-perfect results in a \textit{Same} scenario without meaningful clinical reasoning.
 
\textbf{\textit{Diff}}: 
Our models’ primary advantage appears when the patient’s status changes between the previous and target CXRs.
In these cases, both of our models outperform all baselines, achieving around 80\% AUROC compared to GT performance. 
This result underscores their ability to capture changing pathologies driven by the given medical history.
By contrast, the table classifier variants struggle significantly, particularly the \textit{Table classifier (w/ prev label)}. 
Since this method simply replicates the previous label, it fails to detect changes in pathological findings, leading to notably poor performance on the \textit{Diff} subset.
Overall, performance in the \textit{Diff} subset is generally lower than in \textit{Same}, even for the GT. We hypothesize that this discrepancy arises from rapid status changes (\eg, within two days), which may place patients in a borderline state, making precise predictions more challenging.

\begin{figure}[t!]
\begin{center}
\includegraphics[width=\linewidth]{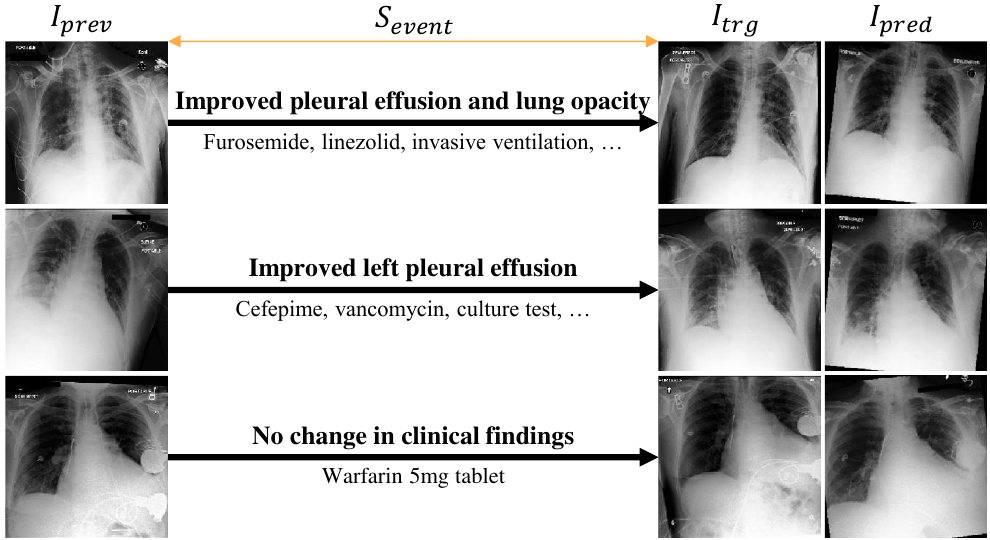}
\end{center}
\vspace{-0.5cm}
\caption{Qualitative results. $I_{prev}$ and $I_{trg}$ are real CXR images from two different timestamps, while $I_{pred}$ is predicted by \mnamenull. $\mathcal{S}_{event}$ (below the arrows) represents the medical events between the two timepoints, with \textbf{Bold} indicating descriptions of the differences between the CXRs.}
\vspace{-0.5cm}
\label{fig:sample}
\end{figure}

\subsection{Preservation of Demographic Information} 
We present the performance of demographic preservation in \tableref{table:test_demo}. 
Since the demographic information of the GT target CXR should remain consistent with the previous image, we see that demographic measurements in the \textit{Previous image} row closely match those of the GT.
In the \mname and \mnamenull rows, we observe that incorporating a ``null'' concept (\mnamenull) improves demographic preservation compared to \mname alone. This suggests that our data augmentation approach effectively teaches the model to preserve structural information not directly related to the medical events. Although gender and race information is successfully maintained, preserving accurate age data remains more challenging. This implies that gender and race cues might be more readily identifiable in CXR images, whereas extracting precise age from such images is inherently more difficult.

\subsection{Image Fidelity and Qualitative Results}
\tableref{table:test_demo} shows that the \mname and \mnamenull models achieve low FID scores (4-6), which are comparable to SOTA text-conditional CXR generation models~\citep{chambon2022roentgen}, reporting scores between 3 and 9. This finding highlights the ability of our models to generate realistic, high-quality CXR images, as illustrated in \figureref{fig:sample}.
The \mnamenull model effectively captures temporal changes, such as resolving the white appearance in the left lung, as observed in both $I_{trg}$ and $I_{pred}$ in rows 1–2, while maintaining the overall status in row 3. However, there are limitations in preserving finer details. For example, the model detects only one device, missing the other in row 3. 
Additional qualitative samples can be found in \sectionref{apd:qualitative_result}.

\begin{table}[t!]
  \caption{Ablation study of different conditioning inputs for our model. Results are categorized into \textit{All} (overall performance), \textit{Same} (identical $I_{prev}$ and $I_{trg}$ labels), and \textit{Diff} (different labels). We report macro-weighted AUROC as mean $\pm$ std across three seeds.}
  \vspace{-0.3cm}
  \begin{center}
    {\small{
    \resizebox{0.99\linewidth}{!}{%
        \begin{tabular}{c|c|ccc|*{3}{>{\centering\arraybackslash}p{1.5cm}}}
            \toprule
            \# & concat & \multicolumn{3}{c|}{cross-attn}                                                          & \multicolumn{3}{c}{Chest ImaGenome}    \\ 
            \midrule            
            & $\textit{E}_{VAE}$ & $\textit{E}_{VAE}$ & $\textit{E}_{CLIP}^{img}$ & $\textit{E}_{CLIP}^{tab}$ &    \textit{All}       &       \textit{Same}       &       \textit{Diff}       \\
            \midrule            
            1  & \checkmark  &            &            &            &  $0.646_{ \pm 0.001}$  &  $0.709_{ \pm 0.002}$ &  $0.536_{ \pm 0.007}$     \\  
            2  &            & \checkmark &            &            &  $0.621_{ \pm 0.001}$  &  $0.674_{ \pm 0.008}$ &  $0.537_{ \pm 0.026}$     \\  
            3  &            &            & \checkmark &            &  $0.746_{ \pm 0.004}$  &  $0.836_{ \pm 0.010}$ &  $0.567_{ \pm 0.007}$     \\  
            4  &            &            &            & \checkmark &  $0.624_{ \pm 0.003}$  &  $0.661_{ \pm 0.010}$ &  $0.549_{ \pm 0.007}$     \\  
            5  & \checkmark  &            &            & \checkmark &  $0.652_{ \pm 0.006}$  &  $0.700_{ \pm 0.010}$ &  $0.564_{ \pm 0.005}$     \\  
            6  & \checkmark  &            & \checkmark & \checkmark &  $0.725_{ \pm 0.001}$  &  $0.797_{ \pm 0.006}$ &  $0.569_{ \pm 0.008}$    \\  
            7  &            & \checkmark & \checkmark & \checkmark &  $0.723_{ \pm 0.003}$  &  $0.796_{ \pm 0.003}$ &  $0.576_{ \pm 0.005}$    \\  
            \bottomrule
        \end{tabular}
    }}}
  \end{center}
  \vspace{-0.5cm}
  \label{table:ablation}
\end{table}


\subsection{Ablation Study}
We analyze the effectiveness of our conditioning inputs by comparing the test set performance in \tableref{table:ablation}. Rows 1-4 illustrate the impact of various conditioning methods. Rows 1 and 2 use the same VAE embedding, but differ in how it is fed into the model: concatenation with the U-Net input or cross-attention via attention layers of U-Net. When using only image conditions (rows 1-3), we observe high performance in the \textit{Same} subset. The CLIP embedding outperforms the VAE embedding due to its high-level features, which are more directly related to CXR pathology. However, this approach has limitations in the \textit{Diff} subset. With only table embeddings (row 4), the model struggles due to the lack of visual context.

In rows 5-6, we concatenate the VAE embedding with the U-Net input while using attention layers to cross-attend to the CLIP embeddings, similar to Instruct-Pix2Pix~\citep{brooks2023instructpix2pix}. In row 7, all features are integrated into the U-Net of LDM using cross-attention. The results indicate that the two different CLIP embeddings provide complementary information (rows 5–6), and combining all features (row 7) further improves performance on the \textit{Diff} subset by effectively leveraging information from multiple sources. Since row 7 demonstrates the best performance in the \textit{Diff} subset, which is important to capturing the temporal state changes, while maintaining comparable performance in the \textit{Same} subset, we select it as our model architecture.

We additionally conducted ablation studies on pretraining, data preprocessing, and variations in conditional medical events to assess our model's effectiveness. Please refer to \sectionref{apd:supp_results} for details.

\section{Discussion}
\paragraph{Limitations} Our framework has been carefully designed, but several limitations remain. 
First, our experiments rely on the MIMIC database, the only open-source resource that includes both image and tabular data, which may limit the generalizability of our findings.
Second, our evaluation depends on the evaluator model, and performance still falls short of the GT, particularly in fine-grained detail preservation. 
Future work should address these issues by expanding data diversity, developing more robust evaluation metrics including human evaluation, and enhancing detail preservation.

\paragraph{Future Direction} Our study marks a significant step in generative models for longitudinal CXR imaging, but further refinement is needed. Future directions include: 1) Extending our framework for long-term prediction by incorporating a retriever mechanism~\citep{kim2023remed}; 2) Integrating additional data types, such as radiology reports; and 3) Evaluating generalization on other datasets or imaging modalities, such as CT. These efforts will lead to more robust and versatile applications.

\section{Conclusion}
In this paper, we introduce a novel task of predicting a future CXR by integrating a prior CXR with subsequent medical events. To address this problem, we propose a latent diffusion-based framework, \mname.
We comprehensively validate \mname's performance, demonstrating its potential to capture temporal changes in CXR findings, as well as its ability to generate high-quality, realistic CXR images. These findings suggest that \mname could be further refined to provide valuable insights for patient monitoring and treatment planning. 
Future research could extend this framework by exploring additional modalities, extending its capabilities to long-term forecasting, and improving detail preservation.

\acks{
This work was supported by the Institute of Information \& Communications Technology Planning \& Evaluation (IITP) grant (No.RS-2019-II190075, No.RS-2022-II220984), and the Korea Health Industry Development Institute (KHIDI) grant (No.HR21C0198), funded by the Korea government (MSIT, MOHW).
}

\bibliography{chil-sample}

\begin{thebibliography}{44}
\providecommand{\natexlab}[1]{#1}
\providecommand{\url}[1]{\texttt{#1}}
\expandafter\ifx\csname urlstyle\endcsname\relax
  \providecommand{\doi}[1]{doi: #1}\else
  \providecommand{\doi}{doi: \begingroup \urlstyle{rm}\Url}\fi

\bibitem[Bae et~al.(2023)Bae, Kyung, Ryu, Cho, Lee, Kweon, Oh, Ji, Chang, Kim, and Choi]{bae2023ehrxqa}
Seongsu Bae, Daeun Kyung, Jaehee Ryu, Eunbyeol Cho, Gyubok Lee, Sunjun Kweon, Jungwoo Oh, Lei Ji, Eric I-Chao Chang, Tackeun Kim, and Edward Choi.
\newblock Ehrxqa: A multi-modal question answering dataset for electronic health records with chest x-ray images.
\newblock In \emph{NeurIPS 2023 Datasets and Benchmarks Track}, 2023.

\bibitem[Bannur et~al.(2024)Bannur, Bouzid, Castro, Schwaighofer, Thieme, Bond-Taylor, Ilse, Pérez-García, Salvatelli, Sharma, Meissen, Ranjit, Srivastav, Gong, Codella, Falck, Oktay, Lungren, Wetscherek, Alvarez-Valle, and Hyland]{bannur2024maira2}
Shruthi Bannur, Kenza Bouzid, Daniel~C. Castro, Anton Schwaighofer, Anja Thieme, Sam Bond-Taylor, Maximilian Ilse, Fernando Pérez-García, Valentina Salvatelli, Harshita Sharma, Felix Meissen, Mercy Ranjit, Shaury Srivastav, Julia Gong, Noel C.~F. Codella, Fabian Falck, Ozan Oktay, Matthew~P. Lungren, Maria~Teodora Wetscherek, Javier Alvarez-Valle, and Stephanie~L. Hyland.
\newblock Maira-2: Grounded radiology report generation, 2024.

\bibitem[Brooks et~al.(2023)Brooks, Holynski, and Efros]{brooks2023instructpix2pix}
Tim Brooks, Aleksander Holynski, and Alexei~A. Efros.
\newblock Instructpix2pix: Learning to follow image editing instructions.
\newblock In \emph{CVPR}, 2023.

\bibitem[Chambon et~al.(2022{\natexlab{a}})Chambon, Bluethgen, Delbrouck, der Sluijs, Połacin, Chaves, Abraham, Purohit, Langlotz, and Chaudhari]{chambon2022roentgen}
Pierre Chambon, Christian Bluethgen, Jean-Benoit Delbrouck, Rogier~Van der Sluijs, Małgorzata Połacin, Juan Manuel~Zambrano Chaves, Tanishq~Mathew Abraham, Shivanshu Purohit, Curtis~P. Langlotz, and Akshay Chaudhari.
\newblock Roentgen: Vision-language foundation model for chest x-ray generation.
\newblock In \emph{arXiv}, 2022{\natexlab{a}}.

\bibitem[Chambon et~al.(2022{\natexlab{b}})Chambon, Bluethgen, Langlotz, and Chaudhari]{chambon2022adapting}
Pierre Joseph~Marcel Chambon, Christian Bluethgen, Curtis Langlotz, and Akshay Chaudhari.
\newblock Adapting pretrained vision-language foundational models to medical imaging domains.
\newblock In \emph{NeurIPS 2022 Foundation Models for Decision Making Workshop}, 2022{\natexlab{b}}.

\bibitem[Cohen et~al.(2020)Cohen, Hashir, Brooks, and Bertrand]{cohen2020limits}
Joseph~Paul Cohen, Mohammad Hashir, Rupert Brooks, and Hadrien Bertrand.
\newblock On the limits of cross-domain generalization in automated x-ray prediction.
\newblock In \emph{Medical Imaging with Deep Learning}, 2020.

\bibitem[Cohen et~al.(2022)Cohen, Viviano, Bertin, Morrison, Torabian, Guarrera, Lungren, Chaudhari, Brooks, Hashir, and Bertrand]{Cohen2022xrv}
Joseph~Paul Cohen, Joseph~D. Viviano, Paul Bertin, Paul Morrison, Parsa Torabian, Matteo Guarrera, Matthew~P Lungren, Akshay Chaudhari, Rupert Brooks, Mohammad Hashir, and Hadrien Bertrand.
\newblock {TorchXRayVision: A library of chest X-ray datasets and models}.
\newblock In \emph{Medical Imaging with Deep Learning}, 2022.

\bibitem[Dasegowda et~al.(2023)Dasegowda, Kalra, Abi-Ghanem, Arru, Bernardo, Saba, Segota, Tabrizi, Viswamitra, Kaviani, Karout, and Dreyer]{Dasegowda2023Suboptimalcxr}
Giridhar Dasegowda, Mannudeep~K. Kalra, Alain~S. Abi-Ghanem, Chiara~D. Arru, Monica Bernardo, Luca Saba, Doris Segota, Zhale Tabrizi, Sanjaya Viswamitra, Parisa Kaviani, Lina Karout, and Keith~J. Dreyer.
\newblock Suboptimal chest radiography and artificial intelligence: The problem and the solution.
\newblock In \emph{Diagnostics}, 2023.

\bibitem[Dosovitskiy et~al.(2021)Dosovitskiy, Beyer, Kolesnikov, Weissenborn, Zhai, Unterthiner, Dehghani, Minderer, Heigold, Gelly, Uszkoreit, and Houlsby]{dosovitskiy2020vit}
Alexey Dosovitskiy, Lucas Beyer, Alexander Kolesnikov, Dirk Weissenborn, Xiaohua Zhai, Thomas Unterthiner, Mostafa Dehghani, Matthias Minderer, Georg Heigold, Sylvain Gelly, Jakob Uszkoreit, and Neil Houlsby.
\newblock An image is worth 16x16 words: Transformers for image recognition at scale.
\newblock In \emph{ICLR}, 2021.

\bibitem[Esser et~al.(2021)Esser, Rombach, and Ommer]{esser2020taming}
Patrick Esser, Robin Rombach, and Björn Ommer.
\newblock Taming transformers for high-resolution image synthesis.
\newblock In \emph{CVPR}, 2021.

\bibitem[Gichoya et~al.(2022)Gichoya, Banerjee, Bhimireddy, Burns, Celi, Chen, Correa, Dullerud, Ghassemi, Huang, Kuo, Lungren, Palmer, Price, Purkayastha, Pyrros, Oakden-Rayner, Okechukwu, Seyyed-Kalantari, Trivedi, Wang, Zaiman, and Zhang]{Gichoya2022race}
Judy~Wawira Gichoya, Imon Banerjee, Ananth~Reddy Bhimireddy, John~L Burns, Leo~Anthony Celi, Li-Ching Chen, Ramon Correa, Natalie Dullerud, Marzyeh Ghassemi, Shih-Cheng Huang, Po-Chih Kuo, Matthew~P Lungren, Lyle~J Palmer, Brandon~J Price, Saptarshi Purkayastha, Ayis~T Pyrros, Lauren Oakden-Rayner, Chima Okechukwu, Laleh Seyyed-Kalantari, Hari Trivedi, Ryan Wang, Zachary Zaiman, and Haoran Zhang.
\newblock Ai recognition of patient race in medical imaging: a modelling study.
\newblock In \emph{Lancet Digit Health}, 2022.

\bibitem[Goodfellow et~al.(2014)Goodfellow, Pouget-Abadie, Mirza, Xu, Warde-Farley, Ozair, Courville, and Bengio]{Goodfellow2014GAN}
Ian Goodfellow, Jean Pouget-Abadie, Mehdi Mirza, Bing Xu, David Warde-Farley, Sherjil Ozair, Aaron Courville, and Yoshua Bengio.
\newblock Generative adversarial nets.
\newblock In \emph{Advances in Neural Information Processing Systems}, 2014.

\bibitem[Gordon(2021)]{cg2021trainCLIP}
Cade Gordon.
\newblock train-clip.
\newblock \url{https://github.com/Zasder3/train-CLIP}, 2021.

\bibitem[Gu et~al.(2023)Gu, Yang, Usuyama, Li, Zhang, Lungren, Gao, and Poon]{gu2023biomedjourney}
Yu~Gu, Jianwei Yang, Naoto Usuyama, Chunyuan Li, Sheng Zhang, Matthew~P. Lungren, Jianfeng Gao, and Hoifung Poon.
\newblock Biomedjourney: Counterfactual biomedical image generation by instruction-learning from multimodal patient journeys.
\newblock In \emph{arXiv}, 2023.

\bibitem[Hayat et~al.(2023)Hayat, Geras, and Shamout]{hayat2023medfuse}
Nasir Hayat, Krzysztof~J. Geras, and Farah~E. Shamout.
\newblock Medfuse: Multi-modal fusion with clinical time-series data and chest x-ray images.
\newblock In \emph{In Proceedings of the 7th Machine Learning for Healthcare Conference}, 2023.

\bibitem[Heusel et~al.(2017)Heusel, Ramsauer, Unterthiner, Nessler, and Hochreiter]{Heusel2017fid}
Martin Heusel, Hubert Ramsauer, Thomas Unterthiner, Bernhard Nessler, and Sepp Hochreiter.
\newblock Gans trained by a two time-scale update rule converge to a local nash equilibrium.
\newblock In \emph{Neural Information Processing Systems}, 2017.

\bibitem[Huang et~al.(2020)Huang, Pareek, Seyyedi, Banerjee, and Lungren]{Huang2020ehrcxr}
Shih-Cheng Huang, Anuj Pareek, Saeed Seyyedi, Imon Banerjee, and Matthew~P Lungren.
\newblock Fusion of medical imaging and electronic health records using deep learning: a systematic review and implementation guidelines.
\newblock In \emph{npj Digital Medicine}, 2020.

\bibitem[Hur et~al.(2024)Hur, Oh, Kim, Kim, Lee, Cho, Moon, Kim, Atallah, and Choi]{Hur2024GenHPF}
Kyunghoon Hur, Jungwoo Oh, Junu Kim, Jiyoun Kim, Min~Jae Lee, Eunbyeol Cho, Seong-Eun Moon, Young-Hak Kim, Louis Atallah, and Edward Choi.
\newblock Genhpf: General healthcare predictive framework for multi-task multi-source learning.
\newblock In \emph{IEEE Journal of Biomedical and Health Informatics}, 2024.

\bibitem[Ieki et~al.(2022)Ieki, Ito, Saji, Kawakami, Nagatomo, Takada, Kariyasu, Machida, Koyama, Yoshida, Kurosawa, Matsunaga, Miyazawa, Ozaki, Onouchi, Katsushika, Matsuoka, Shinohara, Yamaguchi, Kodera, Higashikuni, Fujiu, Akazawa, Iguchi, Isobe, Yoshikawa, and Komuro]{Ieki2022age}
Hirotaka Ieki, Kaoru Ito, Mike Saji, Rei Kawakami, Yuji Nagatomo, Kaori Takada, Toshiya Kariyasu, Haruhiko Machida, Satoshi Koyama, Hiroki Yoshida, Ryo Kurosawa, Hiroshi Matsunaga, Kazuo Miyazawa, Kouichi Ozaki, Yoshihiro Onouchi, Susumu Katsushika, Ryo Matsuoka, Hiroki Shinohara, Toshihiro Yamaguchi, Satoshi Kodera, Yasutomi Higashikuni, Katsuhito Fujiu, Hiroshi Akazawa, Nobuo Iguchi, Mitsuaki Isobe, Tsutomu Yoshikawa, and Issei Komuro.
\newblock Deep learning-based age estimation from chest x-rays indicates cardiovascular prognosis.
\newblock In \emph{Communications Medicine}, 2022.

\bibitem[Irvin et~al.(2019)Irvin, Rajpurkar, Ko, Yu, Ciurea{-}Ilcus, Chute, Marklund, Haghgoo, Ball, Shpanskaya, Seekins, Mong, Halabi, Sandberg, Jones, Larson, Langlotz, Patel, Lungren, and Ng]{irvin2019chexpert}
Jeremy Irvin, Pranav Rajpurkar, Michael Ko, Yifan Yu, Silviana Ciurea{-}Ilcus, Christopher Chute, Henrik Marklund, Behzad Haghgoo, Robyn~L. Ball, Katie~S. Shpanskaya, Jayne Seekins, David~A. Mong, Safwan~S. Halabi, Jesse~K. Sandberg, Ricky Jones, David~B. Larson, Curtis~P. Langlotz, Bhavik~N. Patel, Matthew~P. Lungren, and Andrew~Y. Ng.
\newblock Chexpert: A large chest radiograph dataset with uncertainty labels and expert comparison.
\newblock In \emph{AAAI}, 2019.

\bibitem[Johnson et~al.(2023{\natexlab{a}})Johnson, Bulgarelli, Pollard, Horng, Celi, and Mark]{Johnson2023mimic2.2}
Alistair Johnson, Lucas Bulgarelli, Tom Pollard, Steven Horng, Leo~Anthony Celi, and Roger Mark.
\newblock Mimic-iv, 2023{\natexlab{a}}.

\bibitem[Johnson et~al.(2024)Johnson, Lungren, Peng, Lu, Mark, Berkowitz, and Horng]{Johnson2019mimiccxrjpg}
Alistair Johnson, Matthew Lungren, Yifan Peng, Zhiyong Lu, Roger Mark, Seth Berkowitz, and Steven Horng.
\newblock Mimic-cxr-jpg - chest radiographs with structured labels, 2024.

\bibitem[Johnson et~al.(2019)Johnson, Pollard, Berkowitz, Greenbaum, Lungren, Deng, Mark, and Horng]{johnson2019mimiccxr}
Alistair E.~W. Johnson, Tom~J. Pollard, Seth~J. Berkowitz, Nathaniel~R. Greenbaum, Matthew~P. Lungren, Chih-ying Deng, Roger~G. Mark, and Steven Horng.
\newblock Mimic-cxr, a de-identified publicly available database of chest radiographs with free-text reports.
\newblock \emph{Scientific data}, 2019.

\bibitem[Johnson et~al.(2023{\natexlab{b}})Johnson, Bulgarelli, Shen, Gayles, Shammout, Horng, Pollard, Hao, Moody, Gow, Lehman, Celi, and Mark]{johnson2023mimiciv}
Alistair E.~W. Johnson, Lucas Bulgarelli, Lu~Shen, Alvin Gayles, Ayad Shammout, Steven Horng, Tom~J. Pollard, Sicheng Hao, Benjamin Moody, Brian Gow, Li-wei~H. Lehman, Leo~A. Celi, and Roger~G. Mark.
\newblock Mimic-iv, a freely accessible electronic health record dataset.
\newblock \emph{Scientific data}, 2023{\natexlab{b}}.

\bibitem[Kerchberger et~al.(2020)Kerchberger, Bastarache, Shaver, McNeil, Wang, Zheng, Tapdiya, Wei, and Ware]{Kerchberger2020ehrcxr}
V.E. Kerchberger, J.A. Bastarache, C.M. Shaver, J.B. McNeil, C.T. Wang, N.S. Zheng, A.~Tapdiya, W.-Q. Wei, and L.B. Ware.
\newblock Chest radiograph interpretation is critical for identifying acute respiratory distress syndrome patients from electronic health record data.
\newblock In \emph{American Journal of Respiratory and Critical Care Medicine}, 2020.

\bibitem[Kim et~al.(2024)Kim, Shim, Yang, Im, Lim, Jeong, and Choi]{kim2023remed}
Junu Kim, Chaeeun Shim, Bosco Seong~Kyu Yang, Chami Im, Sung~Yoon Lim, Han-Gil Jeong, and Edward Choi.
\newblock General-purpose retrieval-enhanced medical prediction model using near-infinite history.
\newblock In \emph{MLHC}, 2024.

\bibitem[Lee et~al.(2024{\natexlab{a}})Lee, Lee, Kim, Kim, Kim, Kim, Sunwoo, and Choi]{lee2024ViewXGen}
Hyungyung Lee, Da~Young Lee, Wonjae Kim, Jin-Hwa Kim, Tackeun Kim, Jihang Kim, Leonard Sunwoo, and Edward Choi.
\newblock Vision-language generative model for view-specific chest x-ray generation.
\newblock In \emph{CHIL}, 2024{\natexlab{a}}.

\bibitem[Lee et~al.(2024{\natexlab{b}})Lee, Kim, Chang, and Ye]{lee2024llmcxr}
Suhyeon Lee, Won~Jun Kim, Jinho Chang, and Jong~Chul Ye.
\newblock {LLM}-{CXR}: Instruction-finetuned {LLM} for {CXR} image understanding and generation.
\newblock In \emph{The Twelfth International Conference on Learning Representations}, 2024{\natexlab{b}}.

\bibitem[Loey et~al.(2020)Loey, Smarandache, and M.~Khalifa]{Loey2020covidgan}
Mohamed Loey, Florentin Smarandache, and Nour~Eldeen M.~Khalifa.
\newblock Within the lack of chest covid-19 x-ray dataset: A novel detection model based on gan and deep transfer learning.
\newblock \emph{Symmetry}, 12\penalty0 (4), 2020.

\bibitem[Loshchilov and Hutter(2019)]{loshchilov2019adamw}
Ilya Loshchilov and Frank Hutter.
\newblock Decoupled weight decay regularization.
\newblock In \emph{ICLR}, 2019.

\bibitem[Nicolson et~al.(2024)Nicolson, Dowling, Anderson, and Koopman]{Nicolson2024longitudinalcxr}
Aaron Nicolson, Jason Dowling, Douglas Anderson, and Bevan Koopman.
\newblock Longitudinal data and a semantic similarity reward for chest x-ray report generation.
\newblock \emph{Informatics in Medicine Unlocked}, 50:\penalty0 101585, 2024.
\newblock ISSN 2352-9148.
\newblock \doi{https://doi.org/10.1016/j.imu.2024.101585}.

\bibitem[OpenAI et~al.(2024)]{openai2024gpt4technicalreport}
OpenAI et~al.
\newblock Gpt-4 technical report, 2024.

\bibitem[Paszke et~al.(2019)Paszke, Gross, Massa, Lerer, Bradbury, Chanan, Killeen, Lin, Gimelshein, Antiga, Desmaison, K{\"{o}}pf, Yang, DeVito, Raison, Tejani, Chilamkurthy, Steiner, Fang, Bai, and Chintala]{pytorch}
Adam Paszke, Sam Gross, Francisco Massa, Adam Lerer, James Bradbury, Gregory Chanan, Trevor Killeen, Zeming Lin, Natalia Gimelshein, Luca Antiga, Alban Desmaison, Andreas K{\"{o}}pf, Edward~Z. Yang, Zach DeVito, Martin Raison, Alykhan Tejani, Sasank Chilamkurthy, Benoit Steiner, Lu~Fang, Junjie Bai, and Soumith Chintala.
\newblock Pytorch: An imperative style, high-performance deep learning library.
\newblock In \emph{NeurIPS}, 2019.

\bibitem[Radford et~al.(2021)Radford, Kim, Hallacy, Ramesh, Goh, Agarwal, Sastry, Askell, Mishkin, Clark, et~al.]{radford2021clip}
Alec Radford, Jong~Wook Kim, Chris Hallacy, Aditya Ramesh, Gabriel Goh, Sandhini Agarwal, Girish Sastry, Amanda Askell, Pamela Mishkin, Jack Clark, et~al.
\newblock Learning transferable visual models from natural language supervision.
\newblock In \emph{ICML}, 2021.

\bibitem[Rombach et~al.(2022)Rombach, Blattmann, Lorenz, Esser, and Ommer]{rombach2022ldm}
Robin Rombach, Andreas Blattmann, Dominik Lorenz, Patrick Esser, and Björn Ommer.
\newblock High-resolution image synthesis with latent diffusion models.
\newblock In \emph{CVPR}, 2022.

\bibitem[Sanchez-Pinto et~al.(2018)Sanchez-Pinto, Luo, and Churpek]{Sanchez2018icu}
L.~Nelson Sanchez-Pinto, Yuan Luo, and Matthew~M. Churpek.
\newblock Big data and data science in critical care.
\newblock \emph{Chest}, 2018.

\bibitem[Sharma et~al.(2024)Sharma, Salvatelli, Srivastav, Bouzid, Bannur, Castro, Ilse, Bond-Taylor, Ranjit, Falck, et~al.]{sharma2024mairaseg}
Harshita Sharma, Valentina Salvatelli, Shaury Srivastav, Kenza Bouzid, Shruthi Bannur, Daniel~C Castro, Maximilian Ilse, Sam Bond-Taylor, Mercy~Prasanna Ranjit, Fabian Falck, et~al.
\newblock Maira-seg: Enhancing radiology report generation with segmentation-aware multimodal large language models.
\newblock In \emph{ML4H}, 2024.

\bibitem[Shentu and Al~Moubayed(2024)]{Shentu2024CXRIRGen}
Junjie Shentu and Noura Al~Moubayed.
\newblock Cxr-irgen: An integrated vision and language model for the generation of clinically accurate chest x-ray image-report pairs.
\newblock In \emph{WACV}, 2024.

\bibitem[Sundaram and Hulkund(2021)]{sundaram2021ganaug}
Shobhita Sundaram and Neha Hulkund.
\newblock Gan-based data augmentation for chest x-ray classification.
\newblock In \emph{KDD Applied Data Science for Healthcare Workshop}, 2021.

\bibitem[van~den Oord et~al.(2017)van~den Oord, Vinyals, and kavukcuoglu]{van2017VQGAN}
Aaron van~den Oord, Oriol Vinyals, and koray kavukcuoglu.
\newblock Neural discrete representation learning.
\newblock In \emph{Advances in Neural Information Processing Systems}, 2017.

\bibitem[Weber et~al.(2023)Weber, Ingrisch, Bischl, and Rügamer]{weber2023cheff}
Tobias Weber, Michael Ingrisch, Bernd Bischl, and David Rügamer.
\newblock Cascaded latent diffusion models for high-resolution chest x-ray synthesis.
\newblock In \emph{PAKDD}, 2023.

\bibitem[Wu et~al.(2021)Wu, Agu, Lourentzou, Sharma, Paguio, Yao, Dee, Mitchell, Kashyap, Giovannini, Celi, and Moradi]{wu2chestImaGenome}
Joy~T. Wu, Nkechinyere~N. Agu, Ismini Lourentzou, Arjun Sharma, Joseph~A. Paguio, Jasper~S. Yao, Edward~C. Dee, William Mitchell, Satyananda Kashyap, Andrea Giovannini, Leo~A. Celi, and Mehdi Moradi.
\newblock Chest imagenome dataset for clinical reasoning.
\newblock In \emph{NeurIPS Datasets and Benchmarks Track (Round 2)}, 2021.

\bibitem[Yao et~al.(2024)Yao, Yin, Cheung, Liu, and Qin]{DrFuse2024Yao}
Wenfang Yao, Kejing Yin, William~K. Cheung, Jia Liu, and Jing Qin.
\newblock Drfuse: Learning disentangled representation for clinical multi-modal fusion with missing modality and modal inconsistency.
\newblock \emph{Proceedings of the AAAI Conference on Artificial Intelligence}, 38\penalty0 (15):\penalty0 16416--16424, Mar. 2024.
\newblock \doi{10.1609/aaai.v38i15.29578}.

\bibitem[Zhang et~al.(2024)Zhang, Xu, Usuyama, Xu, Bagga, Tinn, Preston, Rao, Wei, Valluri, Wong, Tupini, Wang, Mazzola, Shukla, Liden, Gao, Crabtree, Piening, Bifulco, Lungren, Naumann, Wang, and Poon]{zhang2024biomedclip}
Sheng Zhang, Yanbo Xu, Naoto Usuyama, Hanwen Xu, Jaspreet Bagga, Robert Tinn, Sam Preston, Rajesh Rao, Mu~Wei, Naveen Valluri, Cliff Wong, Andrea Tupini, Yu~Wang, Matt Mazzola, Swadheen Shukla, Lars Liden, Jianfeng Gao, Angela Crabtree, Brian Piening, Carlo Bifulco, Matthew~P. Lungren, Tristan Naumann, Sheng Wang, and Hoifung Poon.
\newblock A multimodal biomedical foundation model trained from fifteen million image–text pairs.
\newblock \emph{NEJM AI}, 2\penalty0 (1), 2024.
\newblock \doi{10.1056/AIoa2400640}.

\end{thebibliography}

\appendix
\pagebreak 
\startcontents[Appendix]
\printcontents[Appendix]{l}{1}{\section*{Appendix Contents}}

\section{Data PreProcessing}
\label{apd:data_preprocess}
For our research, we use three databases, under the PhysioNet license: MIMIC-IV\footnote{\url{https://physionet.org/content/mimiciv/2.2/}}, MIMIC-CXR-JPG\footnote{\url{https://physionet.org/content/mimic-cxr-jpg/2.0.0/}}, and Chest ImaGenome\footnote{\url{https://physionet.org/content/chest-imagenome/1.0.0/}}. 
Specifically, MIMIC-IV provides structured data from electronic health records (EHRs), MIMIC-CXR-JPG contains the associated chest X-ray (CXR) images, and Chest ImaGenome offers a high-quality annotated subset of MIMIC-CXR. 
We extract the triple ($I_{prev}, \mathcal{S}_{event}, I_{trg}$) from MIMIC-IV and MIMIC-CXR-JPG.
The following subsections describe how we link these databases, select relevant CXR images, filter tabular event data, and preprocess events to construct our dataset.

\subsection{Linking Databases} 
To align data across MIMIC-IV and MIMIC-CXR-JPG, we utilize the \texttt{subject\_id} field, which uniquely identifies each patient in both databases.
Since both datasets use a consistent time-shifting process, we can reliably match a patient’s admission and discharge times in MIMIC-IV with the \texttt{StudyDate} and \texttt{StudyTime} fields in MIMIC-CXR-JPG. This alignment enables us to pinpoint which CXR studies were performed during a particular hospital stay.

\subsection{CXR Image Selection} 
For each hospital stay, we first select frontal view images (\ie, PA and AP) taken between admission and discharge. 
Next, we compile all possible pairs of CXR images from the same hospital stay, applying the following constraints: 
\begin{itemize}
    \item Each pair must consist of images captured at distinct timestamps to avoid multiple images from the same study with identical datetime values.
    \item The time interval between the two images must not exceed two days. 
\end{itemize}

\subsection{Table Event Filtering} 
We include only adult patients (age $>$ 18) in our dataset. To calculate a patient’s age at admission, we use the formula: \texttt{age = (admission\_year - anchor\_year) + anchor\_age}. 
We incorporate data from seven main event tables in MIMIC-IV: lab, chart, input, output, prescription, procedure, and microbiology. 
For each event table, medical experts selected items or categories that are most likely to correlate with observable changes in CXR images.
Next, we select the CXR image pair that has at least one selected medical event recorded between their timestamps, and combine it with the corresponding set of filtered events to form a complete sample. 
\tableref{tab:event_num_statistics} summarizes the overall statistics of the medical events per sample (\ie, the $(I_{prev}, \mathcal{S}_{event}, I_{trg})$ triple), both before and after expert filtering. 
Based on these statistics, we set a maximum threshold of 1024 events per sample to ensure computational efficiency and consistency across samples. 
If the number of events exceeds this threshold, we truncate the event list, retaining only the events closest in time to the target CXR study time. This truncation approach helps preserve the most temporally relevant clinical information. We demonstrate the effectiveness of event filtering in \appendixref{apd:ablation_filtering}.
\begin{table}[t!]
\caption{
Statistics of medical events per sample in our overall dataset. We compare the statistic of number of events per sample, defined as the triple $(I_{prev}, \mathcal{S}_{event}, I_{trg})$, before and after expert filtering. We reports the average, minimum, 25th percentile, median (50th percentile), 75th percentile, and maximum number of events per sample.
}
    \begin{center}
        {\normalsize
            \resizebox{0.9\linewidth}{!}{%
            \begin{tabular}{lcc}
            \toprule
                                     & All events & Selected events \\ \midrule
            Avg. \# of events        & 1,385      & 713             \\
            Min                      & 1          & 1               \\
            25th percentile          & 411        & 212             \\
            Median (50th percentile) & 1,251      & 642             \\
            75th percentile          & 2,096      & 1,070           \\
            Max                      & 6,376      & 4,594           \\
        \bottomrule
        \end{tabular}%
    }}
\end{center}
\label{tab:event_num_statistics}
\end{table}

\subsection{Table Event Preprocessing}
After selecting events that occur between the two CXR timestamps, we convert each tabular entry into a free-text string (\eg, “\textit{table\_name column\_name value}...”) following the approach of \citet{Hur2024GenHPF}. We then embed each of these strings using OpenAI’s \texttt{text-embedding-ada-002} model, which produces a 1536-dimensional vector for each event. The resulting event embedding matrix $\mathcal{S}_{event}$ can thus be represented as a matrix of dimension $n_{event} \times 1536$, where $n_{event} \leq 1024$. 

\section{Implementation Details}
\label{apd:implementation_detail}
\subsection{CXR-EHR CLIP Pretraining}
\label{apd:clip_pretrain}
We begin by modifying the original CLIP architecture to replace its text encoder with a custom table encoder. 
Specifically, our table encoder is a two-layer Transformer encoder with 24 attention heads. 
The input feature dimension is set to 1536, matching the output dimension of the OpenAI embedding model. 
After processing through the Transformer encoder, the resulting tabular representations are projected into a 768-dimensional embedding space via a linear layer, ensuring compatibility with the CLIP feature space. 
For the image encoder, we adopt a ViT-B/32 model configured to handle 256×256 input images. 
This model has an embedding dimension of 512, consisting of 12 layers, a patch size of 32, and 8 attention heads in its internal Transformer blocks. The image encoder weights are initialized from a publicly available pretrained checkpoint\footnote{\url{https://github.com/openai/CLIP}}.
Our implementation follows \citet{cg2021trainCLIP}. 
We pretrain these encoders for up to 100 epochs using a batch size of 1024. Optimization is performed using the AdamW~\citep{loshchilov2019adamw} optimizer with an initial learning rate of 5e-4, $\beta_1=0.9$, $\beta_2=0.98$, and a weight decay of 0.2.

\subsection{Data Augmentation}
\label{apd:data_aug_detail}
We adopt an additional data augmentation method to train \mnamenull, as described in \sectionref{sec:data_aug}. For weak augmentation, we apply random rotation within the range of degrees (-10, +10) and random scaling with scale factors ranging from 0.9 to 1.1. Other settings remain the same as described in \sectionref{sec:implement_detail}.

\subsection{Baselines: Fine-tuning of the \textit{Table classifier}}
\label{apd:tab_cls_finetune}
We adopt the same model architecture as the CLIP table encoder ($E_{\text{CLIP}}^{\text{tab}}$) with a linear head for multi-label classification. 
The model is initialized with pretrained CLIP weights (\appendixref{apd:clip_pretrain}) and fine-tuned for up to 100 epochs with a batch size of 128.
We use the AdamW optimizer with an initial learning rate of 1e-4, weight decay of 0.01, and beta values of 0.9 and 0.999.

For the \textit{Table classifier (w/ prev label)}, we introduce a learnable parameter $\mathbf{V} \in \mathbb{R}^{\text{num\_class} \times 1536}$, which matches the table-event embedding dimension.
For each sample, we select the embeddings corresponding to the classes where the label is 1 and compute their average.  
We then concatenate this resulting vector to the input embeddings as the first token, allowing the model to effectively incorporate prior label information.

\section{Evaluation}
\subsection{Diagnosis Classification}
\label{apd:diagnosis_cls}
To evaluate the preservation of clinical information, we use diagnosis classification based on two sets of labels: Chest ImaGenome and CheXpert. 
\subsubsection{Chest ImaGenome} 
\label{apd:chestimagenome_implementation}
Chest ImaGenome is a high-quality annotated dataset derived from MIMIC-CXR. It consists of 242,072 frontal chest X-ray images with corresponding scene graphs that illustrate the relationships between anatomical locations and their attributes. The dataset includes two subsets: the silver dataset, with automatically generated scene graphs for each chest X-ray image, and the gold dataset, a manually validated and corrected subset reviewed by clinicians, derived from 500 unique patients. The reliability of the silver dataset is supported by a high inter-annotator agreement score of 0.984, as evaluated using 500 reports. 

We train the report-level (\ie, image-level) classifier model, using 10 most frequent Chest ImaGenome labels, each with a prevalence between 0.05 and 0.95. 
The classifier model structure and implementation details follow \citet{bae2023ehrxqa}. 
The backbone of our model is a ViT-S/16, pretrained using the Self-Distillation with No Labels (DINO) method~\cite{}. We add a 3-layer MLP head for each attribute. 
For fine-tuning, we train the model for 100 epochs with a batch size of 512. We use a stochastic gradient descent (SGD) optimizer with learning rate of 1e-3, momentum of 0.9. 

\subsubsection{CheXpert} 
\label{apd:chexpert_implementation}
CheXpert is an open-source rule-based tool designed to extract observations from radiology reports, which includes 14 distinct labels. The MIMIC-CXR-JPG database contains structured labels derived from free-text radiology reports using CheXpert. Each label is assigned one of four possible values: NaN (no value present), 1.0 (mentioned and present), 0.0 (mentioned and absent), or -1.0 (mentioned with uncertainty or ambiguous language). For strict evaluation, we include only labels with definitive values, specifically 1.0 (present) and 0.0 (absent).
As a evaluator model, we use a DenseNet-121 model from the XRV collection~\citep{Cohen2022xrv}\footnote{\url{https://github.com/mlmed/torchxrayvision}}, a SOTA image classifier for CheXpert findings. In line with previous works~\citep{gu2023biomedjourney, chambon2022roentgen}, we evaluate performance on the four findings whose AUROC values exceed 0.75 in our test set, to ensure reliable evaluation: Cardiomegaly, Consolidation, Edema, and Pleural Effusion. We report macro-average AUROC across these diagnoses.

\subsection{Demographic Prediction}
\label{apd:demographic_cls}
We use publicly available pretrained checkpoints from the XRV collection for age and race prediction models. For gender classification, we fine-tune the DenseNet-121 model from the XRV collection, originally trained for CXR diagnosis classification, to perform gender classification. The fine-tuning process utilizes all CXR images from the MIMIC-CXR-JPG dataset, excluding studies from patients included in the validation and test sets. The model is fine-tuned for a maximum of 100 epochs with a batch size of 128. We use the AdamW optimizer, initially set with a learning rate of 0.001, weight decay of 0.01, and beta values of 0.9 and 0.999.

\section{Supplementary Results}
\label{apd:supp_results}
\subsection{Result of Chest ImaGenome Classification}
\tableref{table:supp_chestimagenome} presents detailed per-diagnosis results for each model and baselines.
\begin{table}[t!]
    \caption{
    Quantitative results are grouped into \textit{All}, \textit{Same}, and \textit{Diff} categories using CheXpert labels. The \textit{All} category reflects the overall performance across the entire test set, while the \textit{Same} and \textit{Diff} categories correspond to test subsets where the labels for $I_{prev}$ and $I_{trg}$ are identical or different, respectively. We report the macro-average AUROC. For our proposed model, we report the mean $\pm$ standard deviation across three seeds.
    }
    \begin{center}
    {\small{
    \resizebox{1\linewidth}{!}{%
        \begin{tabular}{l|ccc}
            \toprule
                                             & \textit{All} & \textit{Same} & \textit{Diff} \\
            \midrule
            GT                               &  0.802  &  0.851  &  0.697 \\
            \textit{Previous image}                   &  0.768  &  0.846  &  0.572 \\  
            \textit{Previous label}                   &  0.633  &  1.000  &  0.000 \\       
            \midrule
            \textit{Table classifier}                 &  0.637  &  0.696  &  0.531 \\   
            \textit{Table classifier (w/ prev label)} &  0.683  &  0.861  &  0.257 \\    
            \midrule  
            EHRXDiff                         & $0.677_{ \pm 0.016}$ & $0.723_{ \pm 0.009}$ & $0.572_{ \pm 0.072}$ \\  
            $\text{EHRXDiff}_{w\_null}$      & $0.722_{ \pm 0.012}$ & $0.779_{ \pm 0.008}$ & $0.579_{ \pm 0.006}$ \\  
            \bottomrule
        \end{tabular}
    }}}
    \end{center}
    \label{table:main_chexpert}
    \footnotesize $^\ast$ The \textit{Previous image} and \textit{Previous label} baselines both serve as upper bounds for the \textit{Same} subset and as lower bounds for the \textit{Diff} subset.
\end{table}

\begin{table*}[t!]
    \caption{
    Quantitative results on the test set are grouped into \textit{all}, \textit{same}, and \textit{diff} categories, based on Chest ImaGenome labels. 
    For our proposed model, we report the mean and standard deviation of the AUROC score for each label, computed across 3 different seeds (mean $\pm$ std). Performance was evaluated for the following radiological findings: Consolidation (CON), Enlarged Cardiac Silhouette (ECS), Enlarged Hilum (EH), Lung Opacity (LO), Mediastinal Widening (MedW), Pleural Effusion (PEf), Pneumonia (PNM), Pneumothorax (PTX), Pulmonary Edema/Hazy Opacity (PED), and Vascular Congestion (VasC).
    }
    \begin{center}
    {\small{
    \resizebox{1\linewidth}{!}{%
        \begin{tabular}{l|cccccccccc}
            \toprule
            & \makecell{CON} 
            & \makecell{ECS} 
            & \makecell{EH} 
            & \makecell{LO} 
            & \makecell{MedW} 
            & \makecell{PEf} 
            & \makecell{PNM} 
            & \makecell{PTX} 
            & \makecell{PED} 
            & \makecell{VasC} \\ 
            \midrule             
            \multicolumn{11}{c}{\textbf{All}} \\ 
            \midrule             
            GT                              & 0.857 & 0.887 & 0.882 & 0.851 & 0.799 & 0.889 & 0.835 & 0.734 & 0.833 & 0.879 \\
            \textit{Previous image}                  & 0.799 & 0.867 & 0.826 & 0.804 & 0.749 & 0.827 & 0.764 & 0.731 & 0.774 & 0.829 \\  
            \textit{Previous label}                  & 0.593 & 0.691 & 0.563 & 0.677 & 0.549 & 0.699 & 0.590 & 0.746 & 0.647 & 0.548 \\  
            \midrule
            \textit{Table classifier}                & 0.643 & 0.627 & 0.599 & 0.538 & 0.533 & 0.606 & 0.591 & 0.791 & 0.653 & 0.525 \\  
            \textit{Table classifier (w/ prev label)}& 0.713 & 0.739 & 0.613 & 0.752 & 0.538 & 0.748 & 0.645 & 0.872 & 0.675 & 0.631 \\  
            \midrule
            EHRXDiff                        & $0.679_{ \pm 0.013}$ & $0.774_{ \pm 0.009}$ & $0.692_{ \pm 0.017}$ & $0.760_{ \pm 0.015}$ & $0.693_{ \pm 0.014}$ & $0.739_{ \pm 0.002}$ & $0.681_{ \pm 0.007}$ & $0.637_{ \pm 0.017}$ & $0.715_{ \pm 0.010}$ & $0.756_{ \pm 0.009}$ \\  
            $\text{EHRXDiff}_{w\_null}$     & $ 0.717_{ \pm 0.010}$ & $ 0.814_{ \pm 0.001}$ & $ 0.742_{ \pm 0.005}$ & $ 0.800_{ \pm 0.007}$ & $ 0.715_{ \pm 0.018}$ & $ 0.790_{ \pm 0.005}$ & $ 0.704_{ \pm 0.008}$ & $ 0.698_{ \pm 0.007}$ & $ 0.729_{ \pm 0.008}$ & $ 0.803_{ \pm 0.011}$ \\  
            \midrule
            \multicolumn{11}{c}{\textbf{Same}} \\ 
            \midrule
            GT                              & 0.889 & 0.933 & 0.894 & 0.969 & 0.942 & 0.938 & 0.859 & 0.775 & 0.878 & 0.875 \\  
            \textit{Previous image}                  & 0.866 & 0.933 & 0.834 & 0.959 & 0.934 & 0.942 & 0.867 & 0.775 & 0.889 & 0.873 \\  
            \textit{Previous label}                  & 1.000 & 1.000 & 1.000 & 1.000 & 1.000 & 1.000 & 1.000 & 1.000 & 1.000 & 1.000 \\  
            \midrule
            \textit{Table classifier}                & 0.660 & 0.687 & 0.689 & 0.602 & 0.610 & 0.664 & 0.643 & 0.822 & 0.730 & 0.497 \\  
            \textit{Table classifier (w/ prev label)}& 0.951 & 0.999 & 1.000 & 0.990 & 0.997 & 0.994 & 0.927 & 0.978 & 0.971 & 1.000 \\  
            \midrule
            EHRXDiff                        & $0.701_{ \pm 0.019}$ & $0.842_{ \pm 0.003}$ & $0.584_{ \pm 0.042}$ & $0.891_{ \pm 0.008}$ & $0.834_{ \pm 0.010}$ & $0.808_{ \pm 0.002}$ & $0.734_{ \pm 0.013}$ & $0.641_{ \pm 0.017}$ & $0.785_{ \pm 0.011}$ & $0.782_{ \pm 0.011}$ \\  
            $\text{EHRXDiff}_{w\_null}$     & $0.745_{ \pm 0.012}$ & $0.881_{ \pm 0.003}$ & $0.719_{ \pm 0.063}$ & $0.921_{ \pm 0.007}$ & $0.874_{ \pm 0.023}$ & $0.878_{ \pm 0.006}$ & $0.766_{ \pm 0.012}$ & $0.710_{ \pm 0.008}$ & $0.818_{ \pm 0.010}$ & $0.835_{ \pm 0.005}$ \\  
            \midrule
            \multicolumn{11}{c}{\textbf{Diff}} \\ 
            \midrule
            GT                              & 0.751 & 0.726 & 0.778 & 0.633 & 0.633 & 0.749 & 0.787 & 0.544 & 0.734 & 0.819 \\  
            \textit{Previous image}                  & 0.633 & 0.634 & 0.737 & 0.490 & 0.552 & 0.484 & 0.580 & 0.486 & 0.514 & 0.697 \\  
            \textit{Previous label}                  & 0.000 & 0.000 & 0.000 & 0.000 & 0.000 & 0.000 & 0.000 & 0.000 & 0.000 & 0.000 \\  
            \midrule
            \textit{Table classifier}                & 0.652 & 0.492 & 0.447 & 0.432 & 0.381 & 0.465 & 0.480 & 0.597 & 0.480 & 0.598 \\  
            \textit{Table classifier (w/ prev label)}& 0.335 & 0.005 & 0.000 & 0.035 & 0.000 & 0.146 & 0.196 & 0.242 & 0.060 & 0.044 \\  
            \midrule
            EHRXDiff                        & $0.629_{ \pm 0.067}$ & $0.564_{ \pm 0.022}$ & $0.572_{ \pm 0.118}$ & $0.549_{ \pm 0.040}$ & $0.561_{ \pm 0.037}$ & $0.566_{ \pm 0.005}$ & $0.603_{ \pm 0.015}$ & $0.529_{ \pm 0.010}$ & $0.556_{ \pm 0.025}$ & $0.651_{ \pm 0.030}$ \\  
            $\text{EHRXDiff}_{w\_null}$     & $0.651_{ \pm 0.033}$ & $0.606_{ \pm 0.010}$ & $0.552_{ \pm 0.052}$ & $0.578_{ \pm 0.011}$ & $0.452_{ \pm 0.015}$ & $0.549_{ \pm 0.011}$ & $0.611_{ \pm 0.017}$ & $0.550_{ \pm 0.001}$ & $0.538_{ \pm 0.016}$ & $0.647_{ \pm 0.035}$ \\  
            \bottomrule
        \end{tabular}
    }}}
    \end{center}
    \label{table:supp_chestimagenome}
    \vspace{-0.2cm}
    \footnotesize $^\ast$ The \textit{Previous image} and \textit{Previous label} baselines both serve as upper bounds for the \textit{Same} subset and as lower bounds for the \textit{Diff} subset.
\end{table*}

\begin{table*}[t!]
    \caption{
    Quantitative results grouped by \textit{All}, \textit{Same}, and \textit{Diff} categories, using CheXpert labels. The \textit{All} category reflects overall performance across the entire test set, while the \textit{Same} and \textit{Diff} categories correspond to test subsets where the labels for $I_{prev}$ and $I_{trg}$ are identical or different, respectively.
    Results are reported as the average score and standard deviation across three seeds (mean $\pm$ std). Performance is evaluated across four conditions: Cardiomegaly (Ca), Consolidation (CON), Edema (Ed), and Pleural Effusion (PEf).} 
    \begin{center}
    {\small{
    \resizebox{1\linewidth}{!}{%
        \begin{tabular}{l|cccc|cccc|cccc}
            \toprule
            & \multicolumn{4}{c|}{\textit{All}}  
            & \multicolumn{4}{c|}{\textit{Same}} 
            & \multicolumn{4}{c}{\textit{Diff}}\\ 
            \cmidrule(lr){2-13}
            & \makecell{Ca} 
            & \makecell{CON} 
            & \makecell{Ed} 
            & \makecell{PEf} 
            & \makecell{Ca} 
            & \makecell{CON} 
            & \makecell{Ed} 
            & \makecell{PEf} 
            & \makecell{Ca} 
            & \makecell{CON} 
            & \makecell{Ed} 
            & \makecell{PEf} \\ 
            \midrule
            GT                               & 0.793 & 0.816 & 0.786 & 0.814  
                                             & 0.847 & 0.860 & 0.834 & 0.863  
                                             & 0.725 & 0.714 & 0.660 & 0.690  \\
            \textit{Previous image}          & 0.792 & 0.776 & 0.750 & 0.755  
                                             & 0.849 & 0.846 & 0.837 & 0.850  
                                             & 0.722 & 0.535 & 0.523 & 0.506  \\  
            \textit{Previous label}          & 0.612 & 0.577 & 0.654 & 0.688 
                                             & 1.000 & 1.000 & 1.000 & 1.000  
                                             & 0.000 & 0.000 & 0.000 & 0.000  \\       
            \midrule
            \textit{Table classifier}        & 0.626 & 0.677 & 0.650 & 0.595 
                                             & 0.688 & 0.752 & 0.723 & 0.622  
                                             & 0.529 & 0.580 & 0.466 & 0.548  \\   
            \textit{Table classifier (w/ prev label)} & 0.653 & 0.706 & 0.700 & 0.671 
                                             & 0.810 & 0.887 & 0.879 & 0.866  
                                             & 0.421 & 0.118 & 0.263 & 0.225  \\    
            \midrule  
            EHRXDiff                         & $0.701_{ \pm 0.006}$ & $0.628_{ \pm 0.043}$ & $0.707_{ \pm 0.010}$ & $0.674_{ \pm 0.006}$ 
                                             & $0.746_{ \pm 0.004}$ & $0.655_{ \pm 0.036}$ & $0.773_{ \pm 0.003}$ & $0.717_{ \pm 0.008}$ 
                                             & $0.608_{ \pm 0.041}$ & $0.578_{ \pm 0.234}$ & $0.523_{ \pm 0.032}$ & $0.580_{ \pm 0.017}$ \\  
            $\text{EHRXDiff}_{w\_null}$      & $0.725_{ \pm 0.018}$ & $0.703_{ \pm 0.015}$ & $0.721_{ \pm 0.011}$ & $0.740_{ \pm 0.002}$ 
                                             & $0.777_{ \pm 0.021}$ & $0.733_{ \pm 0.015}$ & $0.801_{ \pm 0.019}$ & $0.806_{ \pm 0.001}$ 
                                             & $0.640_{ \pm 0.025}$ & $0.594_{ \pm 0.021}$ & $0.505_{ \pm 0.015}$ & $0.576_{ \pm 0.014}$ \\  
            \bottomrule
        \end{tabular}
    }}}
    \end{center}
    \label{table:supp_chexpert}
    \vspace{-0.2cm}
    \footnotesize $^\ast$ The \textit{Previous image} and \textit{Previous label} baselines both serve as upper bounds for the \textit{Same} subset and as lower bounds for the \textit{Diff} subset.
\end{table*}

\subsection{Results of CheXpert Classification}
\label{apd:chexpert_result}
We report results on CheXpert labels, which include four radiological conditions: Cardiomegaly (Ca), Consolidation (CON), Edema (Ed), and Pleural Effusion (PEf), with selection criteria detailed in \appendixref{apd:chexpert_implementation}. Overall, the trends observed align with those from the Chest ImaGenome labels.
As shown in \tableref{table:main_chexpert}, \mname and \mnamenull consistently outperform the \textit{Previous image} and \textit{Previous label} baselines in both the \textit{All} and \textit{Diff} subsets. Additionally, they achieve performance comparable to \textit{Previous image} (an upper bound) in the \textit{Same} subset. Again, \mnamenull surpasses \mname in overall AUROC (\eg, 0.722 vs. 0.677 for \textit{All}), demonstrating the benefits of null-based augmentation in distinguishing stable findings from changing pathologies. 
The robust performance of our models across both \textit{Same} and \textit{Diff} subsets highlights their ability to leverage information from prior CXR images and corresponding medical history, rather than relying on simplistic shortcuts (\eg, replicating the initial status). We provide the detailed per-diagnosis results for each model and baselines in \tableref{table:supp_chexpert}.

\subsection{Supplementary Evaluation Metrics}
\begin{table}[t!]
  \caption{
  Quantitative results for image quality on the test set (mean $\pm$ std). Img-Img means BioMedCLIPScore between GT CXR image and predicted CXR, and Img-Txt means BioMedCLIPScore between GT CXR repors and predicted CXR.}
  \vspace{-0.5cm}
  \begin{center}
    {\small{
    \resizebox{0.75\linewidth}{!}{%
        \begin{tabular}{l|c|ccc}
            \toprule
                                     &       Img-Img     &       Img-Txt      \\
             \midrule            
             GT                      &      $1.000$               & $0.397$     \\  
             \textit{Previous image} &      $0.903$               & $0.392$     \\  
             \midrule
             \mnamenull              &      $0.888_{ \pm 0.000}$  & $0.384_{ \pm 0.000}$     \\  
            \bottomrule
        \end{tabular}
    }}}
  \end{center}
  \vspace{-0.4cm}
  \label{table:biomedclipscore}
\end{table}
 
We evaluate our model's performance using two additional metrics: Img-Img and Img-Txt BioMedCLIPScore, which measure the similarity of embeddings derived from the BioMedCLIP~\citep{zhang2024biomedclip}. To compute the BioMedCLIPScore, we first calculate the cosine similarity between the GT CXR image embedding and the predicted image embedding (Img-Img). We then compute the similarity between the predicted image embedding and the corresponding GT CXR report (impression section) for the Img-Txt score. 
Both embeddings are extracted using the official BioMedCLIP checkpoint\footnote{\url{https://huggingface.co/microsoft/BiomedCLIP-PubMedBERT_256-vit_base_patch16_224}}. As shown in \tableref{table:biomedclipscore}, our model achieves high scores on both metrics. Note that the previous image baseline serves as an upper bound, since approximately 70\% of the test set features the previous image sharing the same labels as the target (GT) image.

\begin{table}[t!]
  \caption{
  Ablation study on the use of pre-trained CLIP encoders in our model. Results are categorized into \textit{All} (overall performance), \textit{Same} (identical $I_{prev}$ and $I_{trg}$ labels), and \textit{Diff} (different labels). We report macro-weighted AUROC as mean $\pm$ std across three seeds.}
  \vspace{-0.5cm}
  \begin{center}
    {\small{
    \resizebox{0.9\linewidth}{!}{%
        \begin{tabular}{l|ccc}
            \toprule
             \textbf{Pre-trained Encoder}    &    \textit{All}       &       \textit{Same}       &       \textit{Diff}       \\
            \midrule            
            $\times$                    & $0.664 \pm 0.001$ & $0.714 \pm 0.003$ & $0.565 \pm 0.005$ \\
            $\circ$ (\mname)            & $0.723 \pm 0.003$ & $0.796 \pm 0.003$ & $0.576 \pm 0.005$ \\
            \bottomrule
        \end{tabular}
    }}}
  \end{center}
  \vspace{-0.5cm}
  \label{table:pretrain_ablation}
\end{table}

\subsection{Ablation Study on Pretraining Effectiveness}
We analyze the effectiveness of pretraining for our CLIP encoders. In this ablation study, the image encoder weights are initialized from a publicly available pretrained checkpoint\footnote{\url{https://github.com/openai/CLIP}}, while the table encoder is initialized from scratch. All other settings remain the same as in \mname. As shown in \tableref{table:pretrain_ablation}, pretraining significantly enhances the model’s predictive ability across all groups. This is achieved by enabling better alignment of visual and tabular information. This alignment allows the model to extract richer and more relevant features, leading to more accurate and consistent predictions.

\subsection{Ablation Study of Medical Event Selection}
\label{apd:ablation_filtering}
We analyze the effectiveness of medical event filtering in this ablation study. Specifically, we compare model performance using the same model architecture and hyperparameters as \mname, with the only difference being the dataset, one with event selection and one without.  
As shown in \tableref{tab:event_num_statistics}, applying the same threshold for the number of medical events in both settings would be unfair to the setting without event selection. To address this, we set the maximum number of medical events to 2048 for the unfiltered setting.  
Even though the unfiltered setting utilizes twice as many events as the filtered setting, medical event filtering significantly enhances the model’s predictive ability across all groups, particularly in the \textit{Diff} subset (\tableref{table:filtering_ablation}).  
Since EHRs contain an abundance of events, expert-guided event selection helps the model focus on only the most meaningful events, improving its overall performance.
\begin{table}[t!]
  \caption{Ablation study on the use of medical event selection for data pre-processing. In both cases, we use the same model architecture as \mname, without additional data augmentation. Results are categorized into \textit{All} (overall performance), \textit{Same} (identical $I_{prev}$ and $I_{trg}$ labels), and \textit{Diff} (different labels). We report macro-weighted AUROC as mean ± std across three seeds.}
  \vspace{-0.5cm}
  \begin{center}
    {\small{
    \resizebox{\linewidth}{!}{%
        \begin{tabular}{l|c|ccc}
            \toprule
             Event selection    & Max \# of medical events &    \textit{All}       &       \textit{Same}       &       \textit{Diff}       \\
             \midrule            
             \textbf{$\times$}  &           2048          &  $0.717_{ \pm 0.005}$  &  $0.790_{ \pm 0.001}$ &  $0.538_{ \pm 0.008}$     \\  
             \textbf{$\circ$} (\mname)          &           1024          &  $0.723_{ \pm 0.003}$  &  $0.796_{ \pm 0.003}$ &  $0.576_{ \pm 0.005}$    \\  
            \bottomrule
        \end{tabular}
    }}}
  \end{center}
  \vspace{-0.4cm}
  \label{table:filtering_ablation}
\end{table}

\subsection{Ablation Study of Condition Types}
\label{apd:ablation_condition}
\begin{table}[t!]
  \caption{Quantitative result of different conditioning input for our model on the test set (mean ± std). Each conditioning input is integrated into the model via cross-attention within the attention layers of U-Net.}
  \begin{center}
    {\small{
    \resizebox{\linewidth}{!}{%
        \begin{tabular}{ccc|*{3}{>{\centering\arraybackslash}p{1.5cm}}}
            \toprule
            $\textit{E}_{CLIP}^{tab}$ & $\textit{E}_{CLIP}^{img}$ & $\textit{E}_{CLIP}^{txt}$ &    \textit{All}       &       \textit{Same}       &       \textit{Diff}       \\
            \midrule            
             \checkmark &            &            &  $0.624_{ \pm 0.003}$  &  $0.661_{ \pm 0.010}$ &  $0.549_{ \pm 0.007}$     \\  
                        & \checkmark &            &  $0.746_{ \pm 0.004}$  &  $0.836_{ \pm 0.010}$ &  $0.567_{ \pm 0.007}$     \\  
                        &            & \checkmark &  $0.665_{ \pm 0.003}$  &  $0.774_{ \pm 0.010}$ &  $0.440_{ \pm 0.007}$     \\  
            \bottomrule
        \end{tabular}
    }}}
  \end{center}
  \label{table:ablation_txt}
\end{table}

In our setting, we currently focus on image and tabular data since radiology reports, which are free-text descriptions of prior CXR images, only capture a subset of the information present in the images. While these reports describe the patient’s status as reflected in the corresponding CXR and can aid interpretation, the specific details reported may vary depending on the radiologist’s perspective.

To analyze this, we evaluate the effectiveness of various input modalities, including text-based conditions. Specifically, we apply individual conditional inputs, $E_{CLIP}^{tab}$, $E_{CLIP}^{img}$, and $E_{CLIP}^{text}$, which represent the CLIP embeddings for medical events, the previous image ($I_{prev}$), and the text report for $I_{prev}$, respectively. Each conditioning input is then incorporated into the model through cross-attention mechanisms within the attention layers of U-Net.

As shown in \tableref{table:ablation_txt}, the text-based condition outperforms the tabular modality in the \textit{Same} subset. This is likely because the text reports often include details about the patient’s initial status, closely aligned with the information in the previous CXR, thereby effectively setting an upper bound for performance in this subset. However, text-based input still underperforms compared to using the actual imaging modality, which leverages the full spectrum of information available from the prior CXR. This suggests that radiology reports capture only the aspects of the image that radiologists consider most important, leading to potential information loss.

For the \textit{Diff} subset, the text input yields the lowest performance among all modalities. Like the prior CXR images, the reports focus on describing the initial status and do not provide explicit information about expected changes. Moreover, the extra processing step required to convert the text-based reports into an image output format may reduces performance relative to directly using CXR images.

\begin{table}[t!]
  \caption{
 Ablation study analyzing the impact of missing event-related conditioning inputs on model performance. We randomly drop 5\%, 10\%, and 20\% of events from each sample during inference. Results are categorized into \textit{All} (overall performance), \textit{Same} (identical $I_{prev}$ and $I_{trg}$ labels), and \textit{Diff} (different labels).}
  \vspace{-0.5cm}
  \begin{center}
    {\small{
    \resizebox{0.85\linewidth}{!}{%
        \begin{tabular}{c|ccc}
            \toprule
            Drop ratio (\%) &    \textit{All}       &       \textit{Same}       &       \textit{Diff}       \\
            \midrule            
            w/o drop (\mnamenull)  &  $0.764$  &  $0.843$ &  $0.580$    \\ 
            \midrule            
            5\%                    &  $0.759$  &  $0.845$ &  $0.572$    \\  
            10\%                   &  $0.760$  &  $0.845$ &  $0.570$    \\  
            20\%                   &  $0.761$  &  $0.847$ &  $0.567$    \\  
            \bottomrule
        \end{tabular}
    }}}
  \end{center}
  \vspace{-0.3cm}
  \label{table:ablation_drop_events}
\end{table}
\begin{table}[t!]
  \caption{
  Ablation study evaluating the impact of removing specific event categories on model performance. Each row shows the performance when a particular event table is excluded from each sample during inference. The changes compared to the baseline (All tables used) are shown in parentheses, where \textcolor{blue}{+} indicates an increase and \textcolor{red}{-} indicates a decrease. The percentages next to each removed table indicate the proportion of events that the table contributes on average per sample.}
  \vspace{-0.5cm}
  \begin{center}
    {\small{
    \resizebox{\linewidth}{!}{%
        \begin{tabular}{l|ccc}
            \toprule
            Removed Table &    \textit{All}       &       \textit{Same}       &       \textit{Diff}       \\
            \midrule            
            All tables used (\mnamenull) &  $0.764$  &  $0.843$ &  $0.580$    \\ 
            \midrule            
            - chartevents (74.0\%)       &  $0.769$ \textcolor{blue}{(+0.005)}  &  $0.857$ \textcolor{blue}{(+0.014)}  &  $0.571$ \textcolor{red}{(-0.009)}   \\  
            - inputevents (4.7\%)        &  $0.768$ \textcolor{blue}{(+0.004)}  &  $0.853$ \textcolor{blue}{(+0.010)}  &  $0.578$ \textcolor{red}{(-0.002)}   \\  
            - outputevents (2.1\%)       &  $0.759$ \textcolor{red}{(-0.005)}   &  $0.844$ \textcolor{blue}{(+0.001)}  &  $0.566$ \textcolor{red}{(-0.014)}   \\  
            - labevents (4.4\%)          &  $0.749$ \textcolor{red}{(-0.015)}   &  $0.821$ \textcolor{red}{(-0.022)}   &  $0.571$ \textcolor{red}{(-0.009)}   \\  
            - microbiologyevents (1.3\%) &  $0.749$ \textcolor{red}{(-0.015)}   &  $0.822$ \textcolor{red}{(-0.021)}   &  $0.576$ \textcolor{red}{(-0.004)}   \\  
            - prescriptions (13.2\%)     &  $0.726$ \textcolor{red}{(-0.038)}   &  $0.793$ \textcolor{red}{(-0.050)}   &  $0.559$ \textcolor{red}{(-0.021)}   \\  
            - procedureevents (0.3\%)    &  $0.749$ \textcolor{red}{(-0.015)}   &  $0.824$ \textcolor{red}{(-0.019)}   &  $0.575$ \textcolor{red}{(-0.005)}   \\  
           
            \bottomrule
        \end{tabular}
    }}}
  \end{center}
  \vspace{-0.3cm}
  \label{table:ablation_removed_cat}
\end{table}

Although the text modality  (\ie, radiology reports) as a single condition conveys only a subset of the information available in the corresponding CXR, integrating all three modalities, previous CXR images, corresponding reports, and tabular event data, could lead to further improvements. In this integrated approach, the previous CXR provides comprehensive information on the patient’s initial status, the report highlights key findings, and the tabular event data offers additional context to predict changes. This combination is expected to enhance overall model performance, an approach we plan to explore in future work.

\begin{figure*}[t!]
\begin{center}
\includegraphics[width=\linewidth]{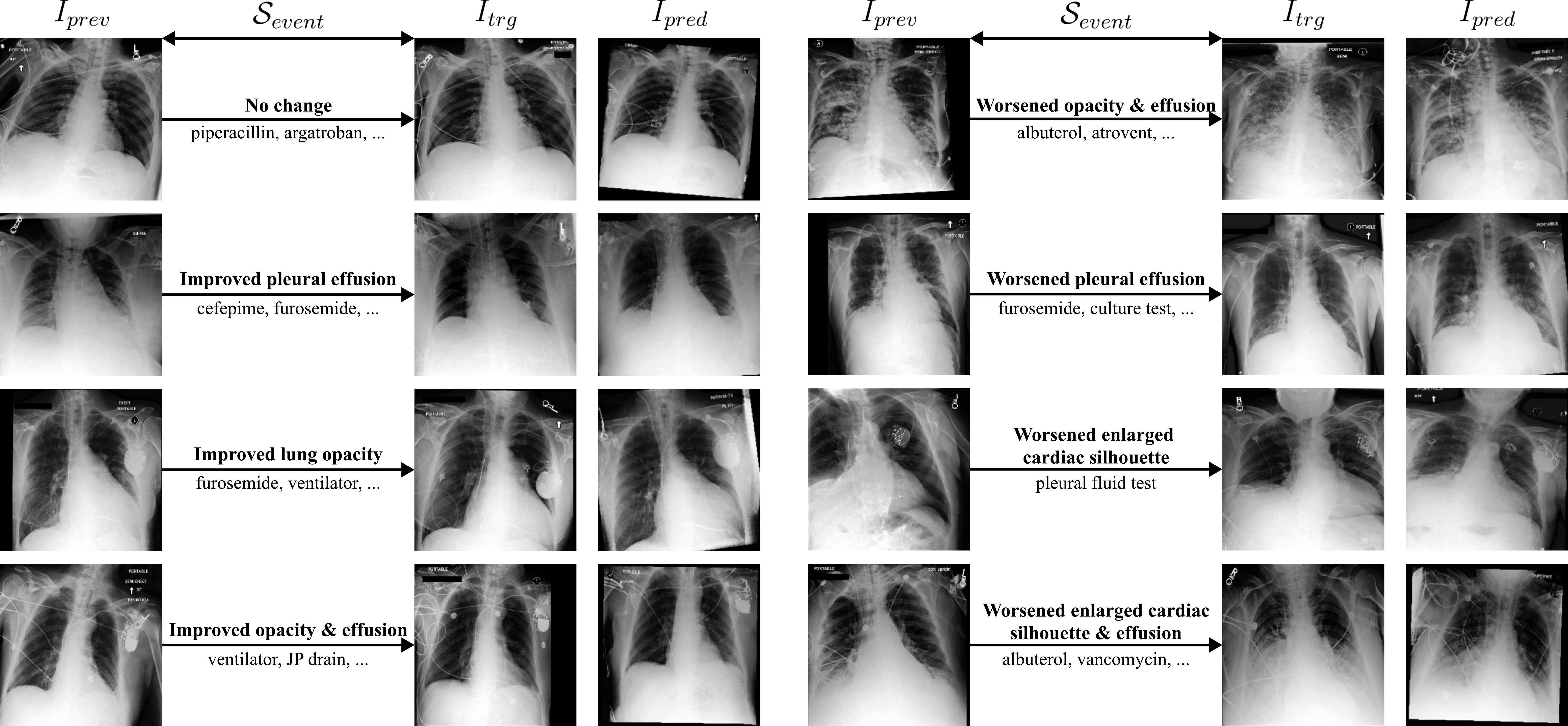}
\end{center}
\vspace{-0.5cm}
\caption{
Qualitative results. $I_{prev}$ and $I_{trg}$ are real CXR images from two different timestamps, while $I_{pred}$ is predicted by \mnamenull. $\mathcal{S}_{event}$ (shown below the arrows) represents the medical events between the two timepoints, with \textbf{Bold} indicating descriptions of the differences between the GT CXRs.}
\vspace{-0.5cm}
\label{fig:additional_sample}
\end{figure*}

\subsection{Ablation Study on Medical Events}
\label{apd:ablation_ehr}
In this section, we analyze the impact of event-level data corruption and evaluate the contribution of each category among our seven core medical event tables (lab, chart, input, output, prescription, procedure, and microbiology). 
These ablation studies are designed to assess our model's sensitivity to noise or missing information and to clarify the influence of individual event types on its performance.
For this ablation study, we use a single seed.

\paragraph{Effect of Randomly Missing Event Data}
We evaluated our model's robustness to variations in EHR data quality during inference by randomly dropping 5\%, 10\%, and 20\% of medical events for each sample. As detailed in \tableref{table:ablation_drop_events}, the model is generally robust to moderate data loss. Compared to the baseline (with no event drop), the overall performance degradation remains marginal even with up to 20\% of events removed.

For the \textit{All} category, the AUROC remains relatively stable, suggesting that the model can effectively handle minor data incompleteness. In the \textit{Same} subset, the AUROC shows little change and even slightly improves. This may indicate that when certain medical events, potentially those suggesting a slight improvement or deterioration in the patient's condition but not significant enough to change the label, are missing, the model becomes more confident in maintaining the original state.

However, in the \textit{Diff} subset, performance declines more noticeably. The AUROC drops from 0.580 in the baseline to 0.567 when 20\% of event data is removed. This suggests that complete event data are crucial for accurately capturing transitions between different clinical states, and missing information increases the ambiguity in tracking disease progression.

Overall, these findings indicate that while the model handles moderate levels of missing EHR data effectively, its performance is more sensitive to missing information when clinical state transitions occur (\textit{Diff}), underscoring the importance of key event data in interpreting changes visible in CXR images.

\paragraph{Effect of Each Category of Medical Events}
To demonstrate the effect of each type of information, we conducted an ablation study by removing specific EHR event categories during inference.
As shown in \tableref{table:ablation_removed_cat}, each event type impacts performance differently based on its role in predicting CXR changes.

Notably, although prescriptions account for only 13\% of the events in each sample, their removal led to the most significant performance drop. Prescriptions include drug orders intended to improve the patient’s condition, and their strong association with CXR changes highlights their clinical importance.

In contrast, while chart events make up the majority of the data, they include redundant information that may be less critical for predicting patient changes compared to other tables. As a result, their removal has a smaller impact on performance than eliminating prescriptions, despite their dominant presence. In the \textit{Same} subset, dropping chart events can even improve performance, as it sometimes removes all events entirely, which is beneficial when no change is expected.

Some medical interventions cause minor fluctuations rather than significant clinical changes, introducing ambiguity into the prediction process. For example, fluid therapy (inputevent) helps maintain a stable state but does not necessarily indicate recovery, while drainage procedures (outputevent) may alleviate conditions like effusion but do not always result in immediate or substantial lung function improvements. Thus, omitting these events biases the model toward predicting no change in borderline cases, sometimes improving performance in the \textit{Same} subset. However, in the \textit{Diff} subset, removing these observations can negatively impact performance.

\subsection{Additional Quantitative Results}
\label{apd:qualitative_result}

\begin{figure}[t!]
\begin{center}
\includegraphics[width=\linewidth]{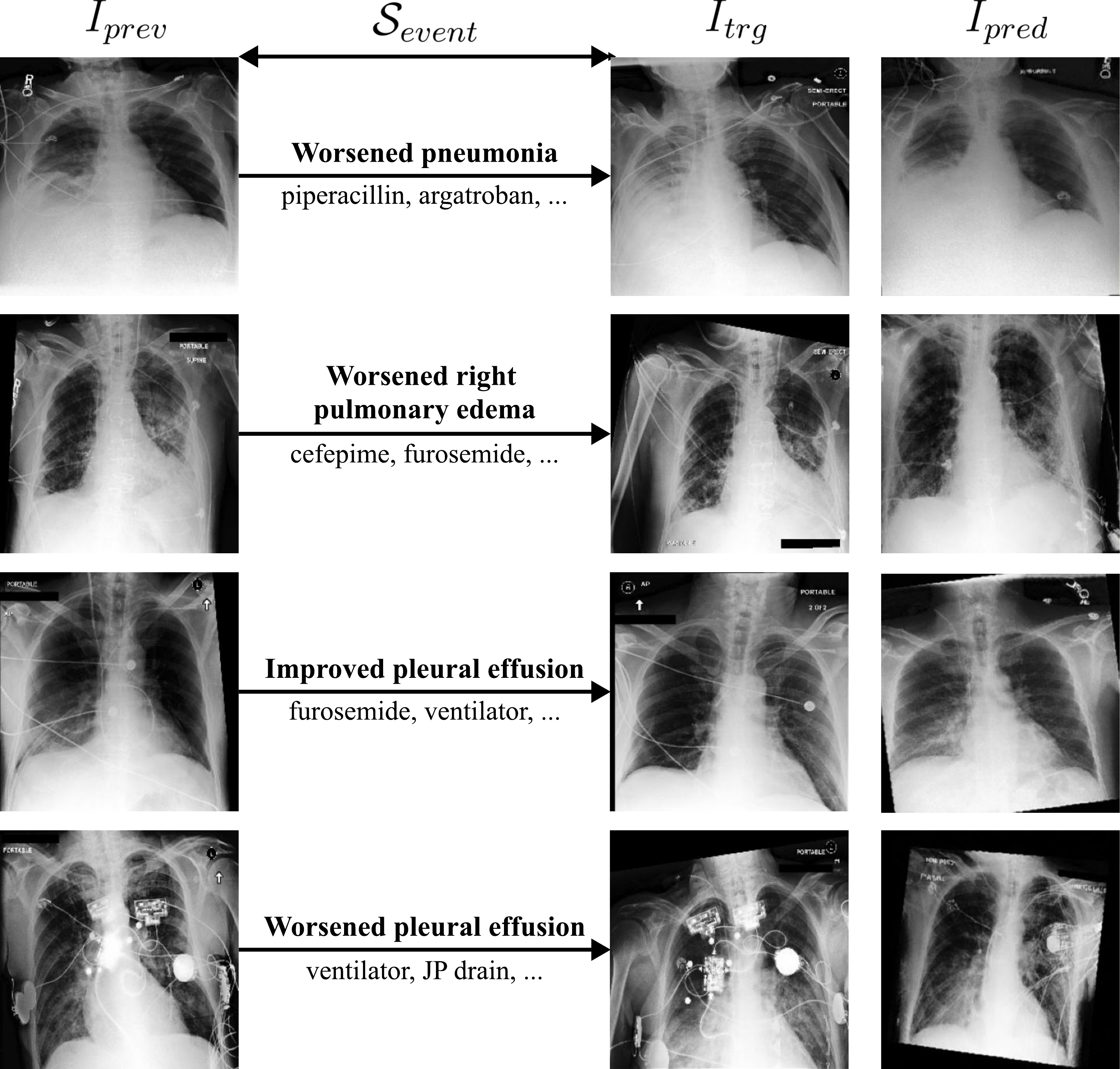}
\end{center}
\vspace{-0.5cm}
\caption{
Qualitative failure case results. $I_{prev}$ and $I_{trg}$ are real CXR images from two different timestamps, while $I_{pred}$ is predicted by \mnamenull. $\mathcal{S}_{event}$ (shown below the arrows) represents the medical events between the two timepoints, with \textbf{Bold} indicating descriptions of the differences between the GT CXRs.}
\vspace{-0.5cm}
\label{fig:failure_sample}
\end{figure}

We provide additional quantitative results in \figureref{fig:additional_sample} and \figureref{fig:failure_sample}. In \figureref{fig:additional_sample}, our model successfully predicts changes in the CXR that align with the GT image ($I_{trg}$). Although there are slight differences in patient positioning due to movement, the clinical findings remain consistent.

However, in some cases, the model fails to generate accurate predictions. For instance, while it correctly captures the overall trend of disease progression, it sometimes misrepresents the severity (rows 1–2 in \figureref{fig:failure_sample}). Another common failure case involves missing or incorrectly rendered medical devices, as shown in rows 3–4. In row 3, the catheter is absent in the predicted image, while in row 4, the model generates a device but misidentifies the type of tube compared to the ground truth.

In cases where the model fails entirely, it may produce unexpected disease progression. For example, in row 3, while the actual patient’s condition shows a cleared right lung, the predicted image incorrectly displays an opacified lung field.

\end{document}